\documentclass[a4paper,11pt]{article}
\usepackage{biblatex}
\usepackage{geometry}
\usepackage{textcomp}
\usepackage{amsmath}
\numberwithin{equation}{section}
\usepackage{amssymb}
\usepackage{graphicx}
\usepackage{color,xcolor}
\usepackage{mathrsfs}
\usepackage{caption}
\usepackage{appendix}
\usepackage[textsize=scriptsize,textwidth=2.5cm]{todonotes}
\usepackage[mathscr]{eucal}
\usepackage[colorlinks=true, allcolors=blue]{hyperref}
\usepackage[hyphenbreaks]{breakurl}
\usepackage{tcolorbox}
\usepackage{tikz}
\usepackage{tikz-cd}
\usepackage{physics}
\usepackage{diagbox}
\usepackage{listings}
\usepackage{array}
\usepackage{subfig}
\usepackage{textcomp}
\usepackage[mathscr]{eucal}
\usepackage[normalem]{ulem}
\def\be{\begin{equation}}
\def\ee{\end{equation}}
\def \ba {\begin{array}}
\def \ea {\end{array}}
\def \bea{\begin{eqnarray}}
\def \eea{\end{eqnarray}}

\def \D {\Delta}

\def \s {\sigma}

\def \nn {\nonumber}

\def\nn{\nonumber}
\def\mD {\mathcal{D}}
  
 \def\cE{{\cal E}}

  \def\cO{{\cal O}}

\def\hs{\hspace}

\newcommand{\ZF}[2]{\textcolor{blue}{#1}\todo[color=green]{\scriptsize{ZF: #2}}}
\begin{document}
\title{ Free field realization of the BMS Ising model}
\vspace{14mm}
\author{
Zhe-fei Yu$^3$ and Bin Chen$^{1,2,3}$\footnote{yuzhefei@pku.edu.cn, bchen01@pku.edu.cn}
}
\date{}

\maketitle

\begin{center}
	{\it
		$^{1}$School of Physics and State Key Laboratory of Nuclear Physics and Technology,\\Peking University, No.5 Yiheyuan Rd, Beijing 100871, P.~R.~China\\
		\vspace{2mm}
		$^{2}$Collaborative Innovation Center of Quantum Matter, No.5 Yiheyuan Rd, Beijing 100871, P.~R.~China\\
		$^{3}$Center for High Energy Physics, Peking University, No.5 Yiheyuan Rd, Beijing 100871, P.~R.~China
	}
	\vspace{10mm}
\end{center}

\begin{abstract}
In this work, we study  the inhomogeneous BMS free fermion theory, and show that it gives a free field realization of the BMS Ising model. We find that besides the BMS symmetry there exists an  anisotropic scaling symmetry in BMS free fermion theory. As a result, the symmetry of the theory gets enhanced to an infinite dimensional symmetry generated by a new type of BMS-Kac-Moody algebra, different from the one found in the BMS free scalar model. Besides the different coupling of the $u(1)$ Kac-Moody current  to the BMS algebra, the Kac-Moody level is nonvanishing now such that the corresponding modules are further enlarged to    BMS-Kac-Moody staggered modules. We show that there exists an underlying $W(2,2,1)$ structure in the operator product expansion of the currents, and the BMS-Kac-Moody staggered modules can be viewed as  highest-weight modules of this $W$-algebra. Moreover we obtain  the BMS Ising model by a fermion-boson duality. This BMS Ising model is  not a minimal model with respect to  BMS$_3$, since the minimal model construction based on BMS Kac determinant always leads to chiral Virasoro minimal models.  Instead, the underlying algebra of the BMS Ising model is the $W(2,2,1)$-algebra, which can be understood as a quantum conformal BMS$_3$ algebra.  %\cb{\sout{Furthermore, (to be contined...) classical version was called conformal BMS$_3$ in  literature.  We find  this  algebra  become `exotic' (c=1) at the quantum level. As a result, the BMS Ising model is the only minimal model with respect to the  BMS-Kac-Moody algebra.}}
%\ZF{}{ Appendix B.2 is completed, please check and revise.}
%\ZF{}{I revise 3 points in the following. 1. Add a citation for the paper of Hao et al. 2.Correct the typo mentioned by Prof. Song. 3. Add a footnote for the T transformation. }

\end{abstract}

\baselineskip 18pt
\newpage

\tableofcontents{}

\newpage

\section{Introduction}
Free field theories are the simplest examples of two dimensional(2d) conformal field theories(CFTs). They are important and useful in the study of 2d CFTs. Firstly, on their own right, they are solvable, and are essential in string theories. Some important ingredients in 2d CFTs, such as state-operator correspondence, operator product expansion (OPE) and modular invanriance, can be shown explicitly in free field theories. Secondly, the free fields can be used to study more generic CFTs, which has free field realizations (see, for example, \cite{DiFrancesco:1997nk}).  Standard examples including the Wakimoto representations of affine current algebras and the vertex operator representations of affine current algebras at level one. In particular, the Coulomb gas formalism \cite{Dotsenko:1984ad}, which is based on free bosons with background charges, is very useful in studying Virasoro minimal models \cite{Belavin:1984vu}. Besides, the simplest minimal model, the 2d Ising model, in fact have another free field realization in terms of  free Majorana fermions. Recently, this fermion-boson duality was generalized to all the minimal models \cite{Hsieh:2020uwb}.

In the past decade, there has been increasing interests in studying the field theories with nonrelativistic conformal symmetries. These symmetries often partially break the Lorentz symmetry but keep some kind of scale invariance. They include Schr\"odinger symmetry, Lifshitz symmetry, Galilean conformal symmetry and Carrollian conformal symmetry etc. Especially in two dimensions, the global scaling symmetries could be enhanced to local ones under suitable conditions, which are generated by infinite dimensional algebra \cite{Polchinski:1987dy,Hofman:2011zj,Chen:2019hbj,Chen:2020juc}. On the other hand, these two dimensional nonrelativistic conformal symmetries play important roles in establishing the holographic dualities beyond the AdS/CFT correspondence \cite{Maldacena:1997re,Brown:1986nw,Strominger:1997eq,Guica:2008mu,Hofman:2011zj,Compere:2013aya,Compere:2013bya,Compere:2009zj,Bagchi:2010zz}.  Among these dualities, the flat-space holography is of particular interest. It was found that the asymptotic symmetry group of three-dimensional Einstein gravity is generated by the Bondi-Metzner-Sachs (BMS) algebra, whose generators satisfy the following commutation relations
\begin{equation}
\begin{split}
&[L_n,L_m]=(n-m)L_{n+m}+\frac{c_L}{12}(n^3-n)\delta_{n+m,0},\\
&[L_n,M_m]=(n-m)M_{n+m}+\frac{c_M}{12}(n^3-n)\delta_{n+m,0},\\
&[M_n,M_m]=0.
\end{split}\label{BMSalgebra}
\end{equation}
This motivates a lot of works establishing holography theory in asymptotic flat spacetimes, see\cite{Bagchi:2012cy,Barnich:2012xq,Bagchi:2012xr,Fareghbal:2013ifa,Bagchi:2014iea,Jiang:2017ecm,Hijano:2018nhq,Hijano:2019qmi,Apolo:2020bld,Apolo:2020qjm}.

The study of the field theory with BMS symmetry, or equivalently the Galilean conformal symmetry, is interesting itself, besides the implication on flat space holography. Firstly the algebra \eqref{BMSalgebra} could be obtained by taking either the Carrollian (ultra-relativistic) limit or the Galilean (non-relativistic) limit of the usual Virasoro algebra. This is in contrast with the situations in higher dimensions, where two limits lead to different algebras \cite{Chen:2021xkw}. Secondly, the field theory with BMS symmetry is typically not unitary and exhibit novel features \cite{Bagchi:2009pe,Chen:2020vvn,Chen:2022jhx,Saha:2022gjw}. Nevertheless, the BMS (GCA) bootstrap seems still viable, at least for generalized free field theories. For more studies on the GCA bootstrap, see \cite{Chen:2020vvn,Chen:2022cpx,Chen:2022jhx,Bagchi:2016geg,Bagchi:2017cpu}\footnote{The bootstrap program in these works is based on the global BMS symmetry.}. Even though we have some understanding of various aspects of BMS field theories mainly from symmetry constraints, it is necessary to study concrete BMS models to obtain better picture. As a first step, one may start from free field theories. In \cite{Hao:2021urq}, a free scalar theory with the BMS symmetry was carefully studied. In the present work, we would like to study another free theory, the inhomogeneous BMS free fermion theory. 

%BMS field theory (BMSFT), or Galilean conformal field theory (GCFT), is a kind of 2 dimensional non-relativistic conformal field theory with  isotropic scaling symmetry:
%\begin{equation}
%    \mathcal{D}: \quad x\to\lambda x, \quad y\to \lambda y
%\end{equation}
%Note that in general,  non-relativistic conformal field theory can have anisotropic scaling symmetry \cite{Chen:2019hbj}.  BMS field theory was studied in the past few years, however, mainly on its kinematics. In fact, unlike 2d CFT, GCFT is poor in concrete examples: the only known GCFT is the free ones, which originate from the study of tensionless string\footnote{Another free theory is the so-called generalized Galilean free field theory (GGFT). However, its symmetry algebra is the global part of the Galilean conformal algebra (GCA).}.}{}

The inhomogeneous BMS free fermion model arise from the tensionless limit of superstring \cite{Bagchi:2016yyf,Bagchi:2017cte}. We discuss both of its NS sector and Ramond sector, the quantization and the spectrum. We show that due to an extra anisotropic scaling symmetry, the BMS symmetry gets enlarged to a symmetry generated by a new type of BMS-Kac-Moody algebra with nonvanishing Kac-Moody level. The enlarged algebra is different from the one appearing in the BMS free scalar model, as now the commutators between the $u(1)$ generators and the superrotations are non-vanishing. To describe all the states in the theory we need to enlarged the module as well. Actually we show that there appears naturally  BMS-Kac-Moody staggered modules. Moreover, by studying the operator product expansion of various currents,  we notice the emergence of a $W(2,2,1)$-algebra  $\mathfrak{S}$, which could be taken as the quantum version of the conformal BMS algebra studied in \cite{Fuentealba:2020zkf}. Interestingly, these BMS-Kac-Moody staggered modules can be taken as  highest-weight modules of $\mathfrak{S}$.%\cb{(to be continued)}

Another important motivation for this work is to search for the minimal models with BMS symmetry. It is well known that the usual 2d minimal models could be read off from the Kac determinant of the Virasoro algebra \cite{Belavin:1984vu}. However, direct computation of the Kac determinant of the BMS algebra shows that the possible minimal models must be chiral. 
Nevertheless, as we will show, there does exist a nontrivial BMS minimal model, the BMS Ising model.  This model cannot be simply obtained by taking the non-relativistic (NR) limit of the usual 2d Ising model, though its partition function is exactly the NR limit of the one of the Ising model. We show that the inhomogeneous BMS free fermion indeed gives a free field realization of this BMS Ising model, which is similar to the usual boson-fermion duality between the 2d Ising model and the free Majorana  fermion. This BMS Ising model is not a minimal model with respect to the BMS$_3$ algebra, instead, its underlying algebra is the above $W(2,2,1)$-algebra $\mathfrak{S}$.
%We discuss modular invariance of its partition function. %(Moreover, we show that there is no other nontrivial BMS minimal model any more.?) }

%\ZF{As its name indicates, }{}it could be obtained simply by taking the  \textcolor{purple}{(ZF: The point about the NR limit of the Ising model is  subtle. Personally, the free field realization help us to see how this NR limit make sense and what is the underlying structure. Naively taking the NR limit of the Ising model, there are some puzzles. Firstly, the spectrum will be 3 BMS primaries: $\mathbf{1}_{\Delta=0,\xi=0}$, $\sigma_{\Delta=\frac{1}{8},\xi=0}$, $\epsilon_{\Delta=1,\xi=0}$. These states, together with their descendents, will not cover all the states we find in the free field realization. In other words, $K$ can not be seen from the NR limit of the Ising model. Secondly, while we can get  fusion rules from the NR limit,  the corresponding null condition can not simply be seen from the NR limit. Of course, these two puzzles are both related to the enlarged symmetry and the enlarged module we find in the free field realization. So we need to be careful about  statements on the NR limit of the Ising model.)}   But to understand its underlying symmetry, it is better to view it in a free field realization. 

\subsection*{Outline of the paper}

The remaining part of this work is organized as follows. In section \ref{reviewBMS}, we give a brief review on the BMS field theories. It includes the symmetry algebra, its representations and the correlation functions. In section \ref{GIsing}, we discuss the inhomogenous BMS free fermion in its NS sector. We show that the symmetry algebra is a new type of BMS-Kac-Moody algebra (with non-vanishing Kac-Moody level), which interestingly includes an anisotropic scaling symmetry. Besides, in  order to cover all  states in the theory, the corresponding modules need to be further  enlarged to  BMS-Kac-Moody staggered modules. These 
modules  in fact can be viewed as  highest weight modules of a $W(2,2,1)$-algebra $\mathfrak{S}$. We also comment on the differences and similarities with the BMS free scalar model. In section \ref{RamondIsing}, we discuss the Ramond sector and  the BMS Ising model. We firstly analyze the zero modes and find the degenerate (two) ground states of the Ramond sector. These states are created by two twist operators $\sigma$ and $\mu$,  which realize the ``spin operator'' in the BMS Ising model. Then we find the fusion rules and calculate the partition function of the theory.  In appendix \ref{Aniscalar}, we show a similar anisotropic scaling symmetry in the BMS free scalar model. In appendix \ref{boo}, we discuss the bottom-up construction of enlarged BMS algebras, including BMS-Kac-Moody algebras and the quantum conformal BMS algebra. In appendix \ref{BMSKac}, we show that the minimal model construction based on the BMS algebra must be chiral.

{\bf Note:} while we were finishing this project, we were aware that the same BMS free fermion model was being studied by another group \cite{Hao:2022xhq}. Their work has many overlaps with the present paper. 
%\ZF{}{I add a citation here to cite their paper.}

\section{Review of BMS field theory}\label{reviewBMS}
In this section, we give a brief review on BMS field theories (BMSFT) in 2d. More detailed discussions can be found  in \cite{Bagchi:2009pe}\cite{Bagchi:2017cpu}\cite{Chen:2020vvn}\cite{Hao:2021urq}.

\subsection{Basics}
A two dimensional BMS field theory (or Galilean conformal field theory) is a non-relativistic conformal field theory, which is invariant under the following BMS transformation\footnote{In this work, we mainly treat BMSFTs defined on the plane with coordinate $x$ and $y$. One can find the cylinder-to-plane mapping and a detailed analysis of the radial quantization for BMSFTs in \cite{Hao:2021urq}.}:
\begin{equation}\label{BMStrans}
    x\rightarrow f(x), \qquad y\rightarrow f'(x)y+g(x).
\end{equation}
The corresponding generators are
\begin{equation}
\begin{aligned}
    l_n&=-x^{n+1}\partial_x-(n+1)yx^n\partial_y, \\
    m_n&=-x^{n+1}\partial_y.
\end{aligned}
\end{equation}
At the quantum level, the commutators  acquire central charges, and we obtain the quantum BMS algebra or the Galilean conformal algebra (GCA):
\begin{equation}
\begin{aligned}
\left[L_n,L_m\right]&=(n-m)L_{n+m}+\frac{c_L}{12} n(n^2-1)\delta_{n+m,0} ,\\
[L_n,M_m]&=(n-m)M_{n+m}+\frac{c_M}{12} n(n^2-1)\delta_{n+m,0} ,\\
[M_n,M_m]&=0.
\end{aligned}
\end{equation}
These $L_n$ and $M_n$ are called super-rotations and super-translations respectively. 
The above BMS$_3$ (or GCA$_2$) algebra  can be obtained by taking the ultra-relativistic  or  non-relativistic contraction on the 2d conformal algebra \cite{Bagchi:2009pe}. 

The OPE among the stress tensor currents are 
\begin{equation}
  \begin{aligned}
    T(x_1,y_1)T(x_2,y_2)\sim &\frac{c_L}{2(x_1-x_2)^4}+\frac{2T(x_2,y_2)}{(x_1-x_2)^2}+ \frac{\partial_xT(x_2,y_2)}{x_1-x_2}-\frac{(y_1-y_2)\partial_yT(x_2,y_2)}{(x_1-x_2)^2},  \\
   &-\frac{2c_M(y_1-y_2)}{(x_1-x_2)^5} -\frac{4(y_1-y_2)M(x_2,y_2)}{(x_1-x_2)^3}\\
   T(x_1,y_1)M(x_2,y_2)\sim &\frac{c_M}{2(x_1-x_2)^4}+\frac{2M(x_2,y_2)}{(x_1-x_2)^2}+\frac{\partial_xM(x_2,y_2)}{x_1-x_2},\\
   M(x_1,y_1)M(x_2,y_2)\sim &0.
  \end{aligned}
\end{equation}
The currents have the following mode-expansions
\begin{equation}
\begin{aligned}
    T(x,y)&=\sum_n L_nx^{-n-2}-\sum_n(n+2)M_nyx^{-n-3},\\
    M(x,y)&=\sum_n M_nx^{-n-2}.
\end{aligned}
\end{equation}

\subsubsection*{Representations}
At early days, it is believed that the 
states in BMSFTs  could be organized into the BMS primaries and their descendants \cite{Bagchi:2009pe}, just as in CFT$_2$. However, the recent work on the BMS free scalar model shows that there are BMS staggered modules \cite{Hao:2021urq}, which include  states which  are not BMS primaries themselves but  also are not descendants of any other BMS primaries. In fact, we will find such  staggered modules in the BMS inhomogeneous free fermion with some novel features. Nevertheless,   BMS highest-weight modules  always appear in BMSFTs\footnote{They appear as submodules of some enlarged BMS modules in the BMS free scalar and the BMS inhomogeneous free fermion.} and will be helpful for us to understand the Hilbert space of these theories. We review  singlet BMS highest weight modules in the following.

For a BMS singlet highest-weight representation, a BMS primary state $|\mathcal{O}\rangle$ is an  eigenstate of  $M_0$. Denote the corresponding BMS primary operator as $\mathcal{O}(x,y)$, when
 located at the origin, it is labeled by the eigenvalues $(\Delta, \xi)$ of $(L_0, M_0)$ ($\mathcal{O}\equiv\mathcal{O}(0,0)$):
\begin{equation}\label{eigen}
[L_0,\cO]=\Delta \cO,\qquad [M_0,\cO]=\xi \cO,
\end{equation}
where $\Delta$ and $\xi$ are referred to as the conformal weight and the boost charge  respectively. It also obeys the highest-weight conditions
\begin{equation}\label{highestw}
[L_n,\cO]=0,\qquad [M_n,\cO]=0, \qquad n>0.
\end{equation}
Using the state-operator correspondence, one can write the eigen-equations and highest-weight conditions for a BMS primary state: \eqref{eigen} and \eqref{highestw} with $\cO$ replaced by $|\cO\rangle$ and  commutators replaced by actions of generators.
Acting $L_{-n},M_{-n}$ with $n>0$ successively on the primaries, we  get their descendants.
The operators at other positions can be obtained by the translation operator $U=e^{x L_{-1}+y M_{-1}}$,
\begin{equation}
\cO(x,y)=U\cO(0,0)U^{-1}.
\end{equation}
Using the Baker-Campbell-Hausdorff (BCH) formula, the transformation law for  primary operators are
\begin{align}
    [L_n,\cO(x,y)]&=(x^{n+1}\partial_x+(n+1)x^ny\partial_y+(n+1)(x^n\Delta+nx^{n-1}y\xi))\cO(x,y),\label{Lntrans}\\
    [M_n,\cO(x,y)]&=(x^{n+1}\partial_y+(n+1)x^n\xi)\cO(x,y),\ \ \forall n\geq-1,\label{Mntrans}
\end{align}
which can be integrated to  get the transformation law
\begin{equation}
    \cO'(x,y)=|f'|^{\D}\,e^{\xi\frac{g'+yf''}{f'}}\,\cO(x',y'),  
\end{equation}
under the finite transformation $x\rightarrow f(x),y\rightarrow f'(x)y+g(x)$.

One can also define the BMS quasi-primaries, which transform covariantly under the global BMS symmetry generated by $L_{0,\pm 1}$ and $M_{0,\pm 1}$. The BMS quasi-primaries are  characterized by $(\Delta,\xi)$ as well, but $\xi$ appears generally in the form of the Jordan blocks, suggesting that there appears naturally the boost charge multiplet \cite{Chen:2020vvn}.
%\footnote{\textcolor{blue}{A simple example is the stress tensor quasi-primaries multiplet\cite{Chen:2020vvn}. We will see in the following that primaries multiplet could also appear naturally when considering an enlarged underlying algebra, see \eqref{doublet0}. }}. 
In the BMS free scalar model, the Hilbert space  can in fact be organized by BMS quasi-primaries and their global descendants \cite{Hao:2021urq}\cite{Chen:2022jhx}. However, this will not be the case for the inhomogeneous BMS free fermion. 

\subsubsection*{Correlation functions}
By requiring the vacuum is invariant under the global BMS symmetry, the correlation functions of quasi-primary operators are well constrained by the Ward identities. For the singlet, the two-point function
and three-point function are
\begin{equation}\label{pt27}
\begin{aligned}
        G_2(x_1,x_2,y_1,y_2) &= d \,\delta_{\Delta_1,\Delta_2}\delta_{\xi_1,\xi_2}|x_{12}|^{-2\Delta_1}e^{-2\xi_1\frac{ y_{12}}{x_{12}}},\\
G_3(x_1,x_2,x_3,y_1,y_2,y_3) &= c_{123}|x_{12}|^{-\Delta_{123}}|x_{23}|^{-\Delta_{231}}|x_{31}|^{-\Delta_{312}}e^{-\xi_{123}\frac{y_{12}}{x_{12}}}e^{-\xi_{312}\frac{y_{31}}{x_{31}}}e^{-\xi_{231}\frac{y_{23}}{x_{23}}},
\end{aligned}
\end{equation}
where $d$ is the normalization factor of the two-point function, $c_{123}$ is the coefficient of three-point function which encodes dynamical information of the BMSFT, and
\begin{equation}
x_{ij}\equiv x_i-x_j,\ \ y_{ij}\equiv y_i-y_j,\ \ \Delta_{ijk}\equiv\Delta_i+\Delta_j-\Delta_k,\ \ \xi_{ijk}\equiv\xi_i+\xi_j-\xi_k.
\end{equation}
The four-point functions of singlet quasi-primary operators can be determined up to an arbitrary function of the cross ratios,
\begin{equation}
G_4=\langle \prod_{i=1}^4\cO_i(x_i,y_i)\rangle =\prod_{i,j}|x_{ij}|^{\sum_{k=1}^4 -\Delta_{ijk}/3}e^{-\frac{y_{ij}}{x_{ij}}\sum_{k=1}^{4}\xi_{ijk}/3}\mathcal{G}(x,y)
\end{equation}
where the indices $i=1,2,3,4$ label the external operators $\cO_i$, $\mathcal{G}(x,y)$ is called the stripped four-point function and $x$ and $y$ are the cross ratios,
\begin{equation}
x\equiv\frac{x_{12}x_{34}}{x_{13}x_{24}}\qquad \frac{y}{x}\equiv\frac{y_{12}}{x_{12}}+\frac{y_{34}}{x_{34}}-\frac{y_{13}}{x_{13}}-\frac{y_{24}}{x_{24}}.
\end{equation}
%\cb{For the multiplet, the discussion is similar.} 

\subsection{Multiplets}
As we have mentioned, BMS quasi-primary multiplet naturally appear in BMSFTs. In fact, one can also find BMS primary multiplet in BMSFTs \cite{Hao:2021urq}. Since we want to see how the BMS symmetry constrains the correlators, we focus on the general case of the correlators of BMS quasi-primary multiplet.

 Similar to the Logarithmic CFT (LCFT), the boost multiplet appears because $M_0$ acts non-diagonally on the quasi-primary states. Generically, $M_0$ acts as follows,
\begin{equation}
    [M_0,\mathbf{O}]=\boldsymbol{\xi}\mathbf{O}
\end{equation}
where $\mathbf{O}$ denotes a set of quasi-primary operators in the theory and $\boldsymbol{\xi}$ is block-diagonalized
\begin{equation}
\boldsymbol{\xi}=
\begin{pmatrix}
\ddots & & & \\
 &\boldsymbol{\xi}_i& & \\
  & & \boldsymbol{\xi}_j& \\
   & & & \ddots\\
\end{pmatrix}
\end{equation}
with
\begin{equation}
\boldsymbol{\xi}_i=
\begin{pmatrix}
 \xi_i& & & \\
  1& \xi_i& & \\
  & \ddots&\ddots &\\
   & & 1& \xi_i\\
\end{pmatrix}_{r\times r}
\end{equation}
being the Jordan block of rank $r$. The quasi-primaries corresponding to a rank-$r$ Jordan block form a multiplet of rank $r$. 

Under the BMS transformations \eqref{BMStrans}, a multiplet $O_a$  of rank $r$ transform as:
\begin{equation}\label{mftran}
    \tilde{O}_{a}(\tilde{x}, \tilde{y})=\sum_{k=0}^{a}\frac{1}{k!}|f'|^{-\Delta}\,\partial_{\xi}^{k}e^{-\xi\frac{g'+yf''}{f'}}\,O_{a-k}(x,y),
\end{equation}
where $a=0,1,...,r-1$ label the $(a+1)$-th operators in the multiplet.
The two-point functions of  the operators in two  multiplets can be written in the following canonical form \cite{Chen:2020vvn}\footnote{Note that our convention has a sign difference with the one in \cite{Chen:2020vvn}.}
\begin{equation}\label{2pt}
\langle \cO_{ia}(x_1,y_1)\cO_{jb}(x_2,y_2)\rangle =\left\{\begin{array}{ll}
0,& \mbox{for $p<0$,}\\
\delta_{ij} d_r\, |x_{12}|^{-2\Delta_i} e^{-2\xi_i\frac{y_{12}}{x_{12}}}\frac{1}{p!}\left(-\frac{2y_{12}}{x_{12}}\right)^p,& \mbox{otherwise,}\end{array} \right.
\end{equation}
where
\begin{equation}
p=a+b+1-r.
\end{equation}
In the above, the indices $i,j$ label the multiplets. When $i=j$, $r\equiv r_i=r_j$ and $a,b$ label the $(a+1)$-th and the $(b+1)$-th operators in the multiplets $\mathcal{O}_i$ and $\mathcal{O}_j$, respectively.  One can then use \eqref{mftran}
to define the out state of a quasi-primaries at infinity, which will be used to calculate the inner product and the Gram matrix,
\begin{equation}\label{outstate}
\langle O_{a}|=\lim\limits_{ y\rightarrow 0 \atop x\rightarrow \infty}\sum\limits_{k=0}^{a}\langle 0|O_{a-k}(x,y) \frac{1}{k!} \partial_{\xi}^k e^{2\xi\frac{y}{x}} x^{2\Delta}.
\end{equation}
From the two-point function \eqref{2pt}, the inner product of the quasi-primary states in a multiplet are 
\bea 
\langle O_a|O_b\rangle =\lim\limits_{x_1\to \infty,x_2 \to 0, \atop  y_{1}\to 0, y_{2}\to 0}\frac{1}{k!} \partial_{\xi}^k e^{2\xi\frac{y_1}{x_1}} x_1^{2\Delta} \sum_{k=0}^{a}\langle O_{a-k}(x_1,y_1) O_{b}(x_2,y_2)\rangle 
= \delta_{a+b,r-1}. \label{norm}
\eea 

The general form of the three-point function is
\begin{equation}
\langle \cO_{ia}\cO_{jb}\cO_{kc}\rangle =ABC_{ijk;abc},
\end{equation}
where
\begin{equation}
\begin{aligned}
A&=\exp(-\xi_{123}\frac{y_{12}}{x_{12}}-\xi_{312}\frac{y_{31}}{x_{31}}-\xi_{231}\frac{y_{23}}{x_{23}}),\\
B&=|x_{12}|^{-\Delta_{123}}|x_{23}|^{-\Delta_{231}}|x_{31}|^{-\Delta_{312}},\\
C_{ijk;abc}&=\sum_{n_1=0}^{a-1}\sum_{n_2=0}^{b-1}\sum_{n_3=0}^{c-1}c_{ijk}^{(n_1n_2n_3)}\frac{(q_i)^{n_1}(q_j)^{n_2}(q_k)^{n_3}}{a!b!c!},\label{ABCcoef}
\end{aligned}
\end{equation}
with
\begin{equation}
q_i=\partial_{\xi_i}\ln A.
\end{equation}
Here $i,j,k$ label the multiplet and $a$, $b$, $c$ label the $(a+1)$-th, $(b+1)$-th, $(c+1)$-th quasi-primaries in the multiplet $\mathcal{O}_i$, $\mathcal{O}_j$, $\mathcal{O}_k$ respectively. $C_{ijk;abc}$ are dynamical three point coefficients which can not be determined by the symmetry.
Note that the above  expression for the three-point function is valid for both $\xi\neq0$ and $\xi=0$. For $\xi=0$, one should take $\xi=0$ at the end of the computation. 

For the stripped four-point function of identical external operators,  when the propagating operators have non-zero boost charge $\xi_r\neq0$, its global block expansion is
\begin{equation} \label{4pointexpansion}
\mathcal{G}(x,y)=\sum_{\cO_r,\xi_r}\frac{1}{d_r}\sum_{p=0}^{r-1}\frac{1}{p!}\sum_{a,b|a+b+p+1=r}c_ac_b\partial_{\xi_r}^pg^{(0)}_{\Delta_r,\xi_r},
\end{equation}
where $g^{(0)}_{\Delta_r,\xi_r}$ is the global BMS(GCA) block for a propagating singlet with $\xi\neq0$
\begin{equation}
    g^{(0)}_{\Delta_r,\xi_r}=2^{2\Delta_p-2}x^{\Delta_r}(1+\sqrt{1-x})^{2-2\Delta_r}e^{\frac{\xi_r y}{x\sqrt{1-x}}}(1-x)^{-1/2}.  \label{singletblock}
\end{equation}
It can be obtained from the Casimir equation. When the propagating operators have zero boost charge $\xi_r=0$, due to the emergent null states, the corresponding block is more complicated, one can find it in \cite{Chen:2022jhx}.  Remarkably, it can be written as a weighted sum of building blocks, which turns out to be the one dimensional conformal block with non-identical external operators \cite{Chen:2022jhx}.

\section{BMS free fermion: NS sector }\label{GIsing}
In this section, we study the  BMS free fermion theory in its NS sector. In fact, one can find two different types of BMS free fermions arising from the tensionless
(super-)string: homogenous and inhomogenous BMS free fermions.%\footnote{\textcolor{blue}{These two theories seems to be  analogs of the   two simplest WCFTs in \cite{Hofman:2014loa}:  Warped Weyl spinor and Warped bc system. So it may be possible to study such kind of free theories in general anisotropic Galilean CFTs \cite{Chen:2019hbj}. } }. 
The homogenous BMS free fermion theory defined on the plane has the following action \cite{Gamboa:1989px}\cite{Bagchi:2016yyf}
\begin{equation}
  \mathcal{L}=\psi_0\partial_y\psi_0+\psi_1\partial_y\psi_1,
\end{equation}
where $\psi_0$ and $\psi_1$ are two fundamental fermions.
One can easily find that this theory has only Virasoro symmetry and is simply a theory of two free chiral fermions. In particular, all operators have no $y$ dependence. Hence, we will focus on the more interesting case: the inhomogenous BMS free fermion. In the following we  simply call it the BMS free fermion theory.

%\ZF{}{agree with the revision here. }

\subsection{Inhomogenous BMS free fermion}
The action for the BMS free fermion is known as the fermionic part of the inhomogeneous tensionless super-string action defined on the plane. It takes the form \cite{Bagchi:2017cte}\cite{Bagchi:2018wsn}:
\begin{equation}
S=\frac{1}{2}\int dxdy (\psi_1\partial_0\psi_0+\psi_0\partial_0\psi_1-\psi_0\partial_1\psi_0).\label{fermionaction}
\end{equation}
%\ZF{}{agree with the revision here. }
Hereinafter we always use the notation that $\partial_0=\partial_y,  \partial_1=\partial_x$.  It is easy to see that this action is invariant under the BMS transformation:
\begin{equation}
    x\to f(x), \qquad y\to f'(x)y+g(x)
\end{equation}
together with the following transformation rule for the fundamental fields:
\begin{equation}
    \psi_0\to f'^{-\frac{1}{2}}\psi_0, \qquad \psi_1\to f'^{-\frac{1}{2}}\left(\psi_1-\frac{g'+yf''}{2f'}\psi_0\right)
\end{equation}
This transformation rule  indicates  that the two fundamental free fermions $\psi_1$ and $\psi_0$ form a rank-2 BMS primary multiplet $(\frac{1}{2}\psi_0,\psi_1)^{\top}$, with dimensions and boost charges (this can be confirmed by their behaviour under the  BMS transformation)
\begin{equation}
  \Delta=\left(\begin{matrix}
   \frac{1}{2} &0  \\
   0& \frac{1}{2}\end{matrix}\right), \qquad
   \xi=\left(\begin{matrix}
   0 &0  \\
   1& 0\end{matrix}\right).
\end{equation}

%\subsubsection{The classical theory}
The equations of motion from the action \eqref{fermionaction} are
\begin{equation}
   \partial_0\psi_0=0, \qquad \partial_0\psi_1=\partial_1\psi_0,
\end{equation}
from which we have the modes expansion:
\bea \label{modesexp}
   \psi_0(x)&=&\sum_n B_n x^{-n-\frac{1}{2}},\nn\\
   \psi_1(x,y)&=&\sum_n A_n x^{-n-\frac{1}{2}}-(n+\frac{1}{2})B_nx^{-n-\frac{3}{2}}y.
\eea
%\ZF{}{agree with the revision here. }
With the help of equations of motion, the stress tensor
\begin{equation}
   T^\mu_\nu=\frac{\partial\mathcal{L}}{\partial(\partial_\mu\Phi)}\partial_\nu\Phi-\mathcal{L}\delta^\mu_\nu
\end{equation}
has the components:
 \begin{equation}
   T^0_0=-\frac{1}{2}\psi_0\partial_1\psi_0, \qquad T^1_1=\frac{1}{2}\psi_0\partial_1\psi_0,
\end{equation}
 \begin{equation}
   T^0_1=-\frac{1}{2}\psi_1\partial_0\psi_1-\frac{1}{2}\psi_0\partial_1\psi_1, \qquad T^1_0=0.
\end{equation}
It is obvious that:
 \begin{equation}
   T^0_0+T^1_1=0,
\end{equation}
so there are two independent components of the stress tensor. We define the stress tensor operators  $T$ and $M$ as:
\bea
   T&\equiv& T^0_1=-\frac{1}{2}(\psi_1\partial_0\psi_1+\psi_0\partial_1\psi_1), \\
   M&\equiv& T^0_0=-\frac{1}{2}\psi_0\partial_1\psi_0,
\eea
%\ZF{}{agree with the revision here. }
which obey the on-shell relation:
\begin{equation}
   \partial_0T=\partial_1M.
\end{equation}
%where we need to use the equations of motion and also remember that $\psi_i, i=1,2$ is fermions so
%\begin{equation}
% (\psi_i)^2=(\partial_j\psi_i)^2=(\partial_k\partial_j\psi_i)^2=...=0
%\end{equation}

 \subsection{Enlarged symmetry and module}
 In this subsection, we further study the symmetries of the BMS free fermion. We show that there actually exist an additional local symmetry which enlarges the BMS symmetry to a bigger space-time symmetry. This enlarged  space-time symmetry turns out to be a new type of BMS-Kac-Moody symmetry: it is generated by an algebra with the commutators between the $u(1)$ generators and  the supertranslations being the supertranslations, thus it is different from the ones obtained from ultra-relativistic contractions. Consequently the module structure is enlarged as well.
 
 \subsubsection{Enlarged symmetry}
The BMS free fermion theory in fact has an enlarged symmetry. From the action \eqref{fermionaction},
 one can find  it is invariant under another dilation symmetry $\mathcal{D}'$, which scales the coordinates as
\begin{equation}
  \mathcal{D}': \quad  x\to x, \qquad y\to \lambda y.\label{D'1}
\end{equation}
Under this scaling transformation, the fundamental fermions has dimension $\frac{1}{2}$ and $-\frac{1}{2}$ respectively
\begin{equation}
  \mathcal{D}': \quad   \psi_0 \to \lambda^{-\frac{1}{2}}\psi_0, \qquad \psi_1 \to
  \lambda^{\frac{1}{2}}\psi_1 \label{D'}
\end{equation}
such that the action is invariant under $\mathcal{D}'$.
Note that this dilation is different from the BMS dilation $\mD$ which is isotropic and requires the scaling dimension of both fundamental fermions to be $\frac{1}{2}$. The corresponding Noether current of the transformation \eqref{D'} reads
\begin{equation}
\begin{aligned}
      J^0_{\mathcal{D}'}&=-\frac{1}{2}\psi_0\partial_1\psi_0y+
      \frac{1}{2}\psi_0\psi_1=M(x)y-\cE(x,y)\\
      J^1_{\mathcal{D}'}&=0
\end{aligned}
\end{equation}
where we have introduced the current
\be
\cE(x,y)\equiv -\frac{1}{2}\psi_0\psi_1.\ee
From the equation of motion, it is clear that $\partial_0^2\cE =0$, which has the solution
\be
\cE(x,y)=\cE^{(0)}(x)+\cE^{(1)}(x)y.
\ee
The current is conserved
\begin{equation}
    \partial_\mu J^\mu_{\mathcal{D}'}=0,
\end{equation}
which indicates that
\begin{equation}
    \partial_0\cE=M,
\end{equation}
i.e. $\cE^{(1)}=M$. Therefore, we have
\be
J^0_{\mathcal{D}'}=-\cE^{(0)}(x).
\ee
In the following we write $J^0_{\mathcal{D}'}$ as $J$.

%In the last equality for $ j^0_{\mathcal{D}'}$ we have used the equations of motion, and $\epsilon^{(0)}(x)$ is the $y$ independent part of $\epsilon(x,y)$: $\epsilon(x,y)=
%\epsilon^{(0)}(x)+\epsilon^{(1)}(x)y$ (from the equations of motion,  so it has this form).
%So it is clear  $j^\mu_{\mathcal{D}'}$ is conserved:this conservation equation  in fact indicates:

From the action, one can easily find another scaling symmetry $\mathcal{D}''$,
    \bea
        \mathcal{D}'': &\quad x\to \lambda x, &\quad y\to y, \\
    & \quad \psi_0\to \psi_0, &\quad \psi_1\to \lambda^{-1} \psi_1.
    \eea
    %\ZF{}{Here is the typo mentioned by Prof. Song: $\lambda\to \lambda^{-1}$.}
    However, this symmetry is not independent and can be decomposed into
      \begin{equation}
        \mathcal{D}''_{\lambda}=\mathcal{D}'_{\lambda^{-1}}\mathcal{D}_{\lambda},
    \end{equation}
    where $\mathcal{D}$ is the usual dilation in the BMS algebra.

   The scaling symmetry \eqref{D'1}\eqref{D'} can in fact be enhanced to  a local symmetry\cite{Polchinski:1987dy,Hofman:2011zj,Chen:2019hbj},
     \begin{equation}\label{cf1}
         \mathcal{C}_f: \qquad x\to x, \quad y\to f(x)y
     \end{equation}
     with the fundamental fields transforming as
     \begin{equation}\label{Cf}
         \mathcal{C}_f: \qquad \psi_0 \to f^{-\frac{1}{2}}\psi_0, \quad \psi_1\to
         f^{\frac{1}{2}}\left(\psi_1-\frac{yf'}{2f}\psi_0\right).
     \end{equation}
     One can easily check the invariance of the action under $\mathcal{C}_f$. When $f(x)=\lambda$, $\mathcal{C}_f$ reduces to $ \mathcal{D}'$. To find the symmetry generators, focusing on a basis of $f(x)$: $x^n$, $n\in\mathbb{N}$. Setting $f(x)=x^n$, we find  the corresponding charges of $\mathcal{C}_f$  are $-\cE^{(0)}_n\equiv J_n$. Thus, $L_n, M_n$ and $J_n$  form the full underlying symmetry algebra of the theory, which we will show in \eqref{TM} and \eqref{JTM}. 
     
     Here we want to find the classical counterpart (that is, without central terms) of this underlying (quantum) symmetry algebra. Denote the generators of the symmetry transformation \eqref{cf1} as $j_n$, then
      \begin{equation}\label{Jvector}
           j_n= -x^ny\partial_y.
      \end{equation}
     together with the generators of the BMS transformations ($l_n$, $m_n$) : 
       \begin{equation}\label{lmvector}
        \begin{aligned}
          l_n&= -x^{n+1}\partial_x-(n+1)x^ny\partial_y, \\
          m_n&= -x^{n+1}\partial_y,
        \end{aligned}
       \end{equation}
      one finds the following commutation relations
 \begin{equation}\label{ClaBMSK}
 \begin{aligned}
     \left[l_n, l_m\right]&=(n-m)l_{n+m},\\
     [l_n, m_m]&=(n-m)m_{n+m},\\
     [m_n, m_m]&=0,\\
     [l_n, j_m]&=-mj_{n+m}\\
     [m_n, j_m]&=-m_{n+m}\\
     [j_n, j_m]&=0.
 \end{aligned}
 \end{equation}
 This is in fact a new type of BMS-Kac-Moody algebra without central extensions. The quantum symmetry algebra
in fact has non-vanishing central terms in $[L_n, L_m]$ and $[J_n,J_m]$, as can be seen in \eqref{TM} and \eqref{JTM}.

We would like to clarify more clearly the structure of this algebra, especially the differences from other known BMS-Kac-Moody algebras. First of all, it is a new type of BMS-Kac-Moody algebra which can not be obtained by a contraction of two (chiral and anti-chiral) usual relativistic Kac-Moody algebras. In the contracted case, the  BMS-Kac-Moody algebra\footnote{It is better not to call such algebras as BMS-Kac-Moody algebras because they are generated by the contracted Kac-Moody current so $T$ and $M$ are not basic generators. Thus it is called ``Galilean affine algebras'' in \cite{Rasmussen:2017eus} and ``non-Lorentzian Kac-Moody algebras'' in \cite{Bagchi:2023dzx}. However, to stress that this algebra is an enlarged BMS algebra, we adopt the name ``BMS-Kac-Moody algebra'' in this paper.  } will have at least two Kac-Moody currents (further, the number of the currents should be even) \cite{Rasmussen:2017eus} (see also \cite{Bagchi:2023dzx}), while our algebra \eqref{ClaBMSK} contains only one Kac-Moody current\footnote{We have classified possible BMS-Kac-Moody algebras with one  Kac-Moody current in the appendix \ref{classifaction}. More precisely, we show that by considering the most general commutation relations between the BMS algebra and  one $u(1)$ Kac-Moody current, the Jacobi identities restrict the enlarged algebra into three types: Type 1 being the one found in \cite{Hao:2021urq}, Type 2 being the one found in the present work, and Type 3 being some singular cases. This proves the consistency of our algebra from another point of view.}. Secondly, the Kac-Moody current couple differently to the BMS part, comparing with the ones from contractions: here the commutator $[m_n, j_m]$ (or its quantum version $[M_n, J_n]$) is proportional to  $m_{n+m}$, while the ones from contractions have $[M_n, J_m]\varpropto J'_{n+m}$ or $0$ ($J'$ is another Kac-Moody current which lie in the same boost doublet as $J$) \cite{Rasmussen:2017eus, Bagchi:2023dzx}. Interestingly, similar commutators could also appear in higher dimensional enlarged Galilean conformal algebras (GCA) \cite{Bagchi:2009my}. In \cite{Bagchi:2009my},  a possible extension of the GCA (recall that in 2d, the GCA is just the BMS algebra studied in this work) is obtained by adding a $so(d)$ Kac-Moody algebra, and in this extended GCA algebra (equation (3.2) in \cite{Bagchi:2009my}) there is a commutator which is of the form $[J,M]\sim M$ but not $[J,M]\sim J'$ or $0$. %\footnote{\textcolor{purple}{As noted in \cite{Bagchi:2009my}, these $J$ and $M$, together with their commutators, are called Coriolis algebra (group) in the literature.}} 
Note that the extended GCA algebra in  \cite{Bagchi:2009my} only exists in $d>2$. In contrast, the BMS-Kac-Moody algebra \eqref{ClaBMSK} is defined in 2d. This is feasible because our vector field realisation of $J$ is different from the one in \cite{Bagchi:2009my} (Eq. (3.4)).  A crucial point about our vector field realisation of $j_n$ in \eqref{Jvector} is that it can not be obtained by a contraction, while in \cite{Bagchi:2009my} it was shown that at least the finite global part of the algebra (Eq. (3.2) in \cite{Bagchi:2009my}) can be obtained by a contraction.  In summary, our result non-trivially extends the enlarged space-time symmetry in \cite{Bagchi:2009my} for $d>2$ to the $d=2$ case. Moreover, we find it is just the symmetry algebra of a real BMS field theory: the BMS free fermion.

 With this anisotropic scaling symmetry (and the enhanced local symmetry $\mathcal{C}_f$) in hand, one may want to know whether this kind of symmetry can be found in other BMSFTs. In fact, the BMS free scalar model  \cite{Hao:2021urq} indeed has such a symmetry, as we show in appendix \ref{Aniscalar}. Moreover, in appendix \ref{boo} we show that BMSFTs with such a symmetry (and no other extra symmetry currents) will always have the central charge $c_M=0$.
%\end{itemize} 

%\subsubsection{Quantization in the }
Now we discuss the canonical quantization in the NS sector. According to the canonical anti-commutation relations of the fundamental fermion fields and their conjugate momentums,  we obtain the anti-commutation relations of the modes defined in \eqref{modesexp}:
\begin{equation}\label{complexfermodes}
 \{A_n,A_m\}=\{B_n,B_m\}=0, \qquad \{A_n,B_m\}=\delta_{n+m,0}.
\end{equation}
Because the inhomogenous   fermions we consider are real fermions on the cylinder \cite{Bagchi:2018wsn}, we have the following Hermition condition for these modes:
\begin{equation}\label{hermitian}
    A_n^\dagger=A_{-n}, \qquad B_n^\dagger=B_{-n}.
\end{equation}
%}{The revised hermitian condition.}
%\ZF{}{agree with the revision here. I also add a footnote 9 for the boundary condition.}
For the NS sector, we must have $n\in\mathbb{N}+\frac{1}{2}$ to satisfy the anti-periodic condition on the cylinder. %\ZF{}{agree with the revision here. }%\footnote{Anti-periodic boundary condition for the theory on the cylinder.  On the plane. it is periodic.}.
The vacuum is defined by:
\begin{equation}
   A_n|0\rangle=B_n|0\rangle=0, \qquad n>0. \label{vac1}
\end{equation}
It leads to the following prescription for fermionic normal ordering:
\begin{equation}
 :X_nX_m:=\left\{\begin{matrix}
   X_nX_m, \quad n<0,  \\
   -X_mX_n,  \quad n>0.\end{matrix}\right.
\end{equation}
where $X=A,B$.
Now the stress tensor operators as well as the current $\cE$ can be written in terms of  the normal-ordered products of the operators
\bea
   T(x,y)&=&-\frac{1}{2}(:\psi_1\partial_0\psi_1+\psi_0\partial_1\psi_1:), \nn\\
   M(x)&=&-\frac{1}{2}:\psi_0\partial_1\psi_0: , \nn\\
  \cE(x,y)&=&-\frac{1}{2}:\psi_0\psi_1:.
\eea
In the following, we will always treat the normal-ordered products of operators so we omit the normal ordered symbols in $T,M, \cE$, and if necessary we use $\hat{N}$ to denote the normal ordering of composite operators. 

The vacuum \eqref{vac1} is in fact the BMS-Kac-Moody highest weight vacuum, satisfying:
\begin{equation}
 L_n|0\rangle=M_m|0\rangle=J_k|0\rangle=0, \qquad n,m\geq-1, \quad k>-1.
\end{equation}
where $L_n$, $M_m$ and $J_k$ are the BMS-Kac-Moody modes and their explicit expressions will be given in \eqref{nmodes}.
The Hilbert space of this theory consist of the following states:
\begin{equation}\label{stateslist}
\begin{matrix}
  \text{$L_0$ eigenvalue}&  \text{states}\\
                        0&  |0\rangle\\
              \frac{1}{2}&  A_{-\frac{1}{2}}|0\rangle, \quad B_{-\frac{1}{2}}|0\rangle\\
                        1&  A_{-\frac{1}{2}}B_{-\frac{1}{2}}|0\rangle\\
              \frac{3}{2}&  A_{-\frac{3}{2}}|0\rangle,\quad  B_{-\frac{3}{2}}|0\rangle\\
                        2&  A_{-\frac{1}{2}}A_{-\frac{3}{2}}|0\rangle,\quad A_{-\frac{1}{2}}B_{-\frac{3}{2}}|0\rangle, \quad B_{-\frac{1}{2}}A_{-\frac{3}{2}}|0\rangle, \quad B_{-\frac{1}{2}}B_{-\frac{3}{2}}|0\rangle\\
                        ...&...
\end{matrix}
\end{equation}
Our next goal is to find the module decomposition of this Hilbert space. Firstly, we want to find the BMS-Kac-Moody primary states, which are defined as:
\begin{equation}
    L_2|\mathcal{O}\rangle=L_1|\mathcal{O}\rangle=M_2|\mathcal{O}\rangle=M_1|\mathcal{O}\rangle=J_n|\mathcal{O}\rangle=0, \qquad n>0
\end{equation}
and are labeled by the conformal dimension $\Delta$, boost charge $\xi$ and the $u(1)$ charge $\varepsilon$:
\begin{equation}
    L_0|\mathcal{O}\rangle=\Delta|\mathcal{O}\rangle,\quad M_0|\mathcal{O}\rangle=\xi|\mathcal{O}\rangle, \quad J_0|\mathcal{O}\rangle=\varepsilon|\mathcal{O}\rangle.
\end{equation}
It turns out that there are four BMS primary states, among which three  are BMS-Kac-Moody primaries. The four BMS primary operators are:
\begin{itemize}
 \item The singlet $\mathbf{1}$ with dimension $\Delta=0$, boost charge $\xi=0$ and charge $\varepsilon=0$, the corresponding state is $|0\rangle$. It is a BMS-Kac-Moody primary.
  \item The doublet $(\frac{1}{2}\psi_0,\psi_1)$ with
  \begin{equation}
  \Delta=\left(\begin{matrix}
   \frac{1}{2} &0  \\
   0& \frac{1}{2}\end{matrix}\right), \qquad
   \xi=\left(\begin{matrix}
   0 &0  \\
   1& 0\end{matrix}\right), \qquad
   \varepsilon=\left(\begin{matrix}
   \frac{1}{2} &0  \\
   0& -\frac{1}{2}\end{matrix}\right).
\end{equation}
  the corresponding states are $A_{-\frac{1}{2}}|0\rangle$ and $B_{-\frac{1}{2}}|0\rangle$.  They are also BMS-Kac-Moody primaries.
  \item The singlet $\cE\equiv -\frac{1}{2}\hat{N}(\psi_0\psi_1)$, with dimension $\Delta=1$, boost charge $\xi=0$ and charge $\varepsilon=0$, the corresponding state is
  $|\cE\rangle=-\frac{1}{2}B_{-\frac{1}{2}}A_{-\frac{1}{2}}|0\rangle$. It is a BMS primary but not a BMS-Kac-Moody primary. This is expected because $\cE$ itself is just the Kac-Moody symmetry current. Note that from the UR limit point of view, this symmetry current is an emergent one because the relativistic counterpart of $\cE$ is $\psi\bar{\psi}$, which is a non-chiral operator and is not a symmetry current in the relativistic free fermion theory. 
\end{itemize}
%\ZF{}{agree with the revision here. }
Using the anti-commutation relation of the modes, we can calculate the correlators of the fundamental fields
\begin{equation}
\begin{aligned}
  \langle\psi_0(x_1,y_1)\psi_0(x_2,y_2)\rangle &=0,\\
  \langle\psi_0(x_1,y_1)\psi_1(x_2,y_2)\rangle &=\frac{1}{x_1-x_2},\\
  \langle\psi_1(x_1,y_1)\psi_1(x_2,y_2)\rangle &=-\frac{y_1-y_2}{(x_1-x_2)^2},
\end{aligned}
\end{equation}
and find that $(\frac{1}{2}\psi_0,\psi_1)$ indeed form a doublet (in the canonical form). We stress that the `energy operator' $\cE$ is a singlet primary. We may  construct  triplet primaries
\begin{equation}
\mathbf{\cE}\sim \left(
\begin{matrix}
   \hat{N}(\psi_0\psi_0) \\
   \hat{N}(\psi_0\psi_1)\\
   \hat{N}(\psi_1\psi_1)
\end{matrix}\right).
\end{equation}
 However, due to the fermionic nature of $\psi_0$ and $\psi_1$, $:\psi_0\psi_0:=:\psi_1\psi_1:=0$,  so we are left with a singlet. Another interesting property of $\cE$ is that its modes expansion has $y$ dependence,
\be
 \cE(x,y)=\sum_n\cE^{(0)}_nx^{-n-1}+\sum_n\cE^{(1)}_nx^{-n-2}y \equiv\sum_n -J_n          x^{-n-1}+\sum_n M_nx^{-n-2}y
 \ee 
 where the explicit expressions of $J_n, M_n$ will be given in \eqref{nmodes}. Moreover, the two-point function of the energy operator depends only on $x$
\begin{equation}\label{corre cE}
  \langle\cE(x_1,y_1)\cE(x_2,y_2)\rangle=\frac{1}{4(x_1-x_2)^2},
\end{equation}
%\ZF{}{agree with the revision here. }
 as a two-point correlator of a singlet should behave.
   Note that this `energy operator' is the same (same dimension, and being a singlet) as the one obtained by taking the non-relativistic limit of the energy operator in the Ising model.

 \subsubsection{Enlarged module}
To organize all the states in \eqref{stateslist} into the representations of the algebra, it turns out that not only the underlying symmetry needs to be enlarged to be a BMS-Kac-Moody algebra, but the modules should also be enlarged. It is easy to see this point from many  novel non-primary states in the spectrum. For example, the state $A_{-\frac{1}{2}}A_{-\frac{3}{2}}|0\rangle$ is not the  descendant of any primary states. In fact, this is similar with the case of BMS free scalar \cite{Hao:2021urq}. As a result, we need an extra  operator
\begin{equation}
 K(x,y)\equiv -\hat{N}(\psi_1\partial_1\psi_1) ,
\end{equation}
with dimension $\Delta=2$ to enlarge  BMS-Kac-Moody highest weight modules. It
 corresponds to the state
\begin{equation}
|K\rangle=-A_{-\frac{1}{2}}A_{-\frac{3}{2}}|0\rangle.
\end{equation}
In fact, it is just the existence of this operator that prevent the decoupling of the $M_n$ (be null) in the enlarged modules. The theory would reduce to a chiral one if the $M_n$'s get decoupled. This kind of module is referred to as the staggered module, which is well-known in the study of logarithmic CFT \cite{Creutzig:2013hma}.  Note that in the BMS free scalar model, such a $K$ operator, together with $T$ and $M$, form  a stress tensor triplet with dimension $\Delta=2$ and boost charge $\xi=0$. The corresponding three states, $|T\rangle$, $|M\rangle$ and $|K\rangle$ lie in the BMS staggered vacuum module in the BMS free scalar. The case here is a little bit different. Before describing the BMS-Kac-Moody staggered modules in the BMS free fermion, we have to study  operators in the stress tensor multiplet in detail. They lie in the  BMS-Kac-Moody staggered vacuum module.

The first step is to find all  the operators in the stress tensor multiplet and write down the operator product expansions (OPEs) among them. In the case at hand, they turn out to be
\begin{equation}\label{compareMK}
 M(x)=-\frac{1}{2}\hat{N}(\psi_0\partial_1\psi_0), \qquad T(x,y)=-\frac{1}{2}\hat{N}(\psi_1\partial_0\psi_1+\psi_0\partial_1\psi_1), \qquad K(x,y)=-\hat{N}(\psi_1\partial_1\psi_1),
\end{equation}
as well as
\begin{equation}
 \partial_1\cE(x,y)=-\frac{1}{2}\hat{N}(\partial_1\psi_0\psi_1+\psi_0\partial_1\psi_1)=-\frac{1}{2}\hat{N}(-\psi_1\partial_0\psi_1+\psi_0\partial_1\psi_1).
\end{equation}
%Note that here we change the normalization of $\epsilon$ to be $\epsilon=-\frac{1}{2}\psi_0\psi_1$ to make our later expressions become  cleaner. Under this normalization,
%\ZF{}{agree with the revision here. }
Note that we have $\partial_0\cE=M$, $\partial_0(\partial_1\cE)=\partial_1M$ and $\partial_0T=\partial_1M$. One can certainly construct another operator with $\Delta=2$: $\hat{N}(\cE\cE)$, which is the normal ordered product of two $\cE$'s. However it is not an independent operator since\footnote{This is in fact the Sugawara construction based on the current $\cE$. From the correlator \eqref{corre cE}, one knows that the Kac-Moody level is $k=\frac{1}{4}$, so $\frac{1}{2k}\hat{N}(\cE\cE)=2\hat{N}(\cE\cE)$.   }
\begin{equation}\label{TcE}
    2\hat{N}(\cE\cE)=T
\end{equation}
One can see this relation either by writing the OPE of $\cE\cE$ up to the constant term, or by showing the corresponding states are identical via the state-operator correspondence.  %which is $N(\cE\cE)$\footnote{\textcolor{blue}{Alternatively, due to the state-operator correspondence, one can verify \eqref{TcE} by showing the correspond states are identical (using \eqref{nmodes}):
%\begin{equation}
%    2|N(\cE\cE)\rangle=2J_{-1}J_{-1}|0\rangle=\frac{1}{2}\sum_m :B_{-1-m}A_m:
%B_{-\frac{1}{2}}A_{-\frac{1}{2}}|0\rangle=\frac{1}{2}(B_{-\frac{3}{2}}A_{\frac{1}{2}}-A_{-\frac{3}{2}}B_{\frac{1}{2}})
%B_{-\frac{1}{2}}A_{-\frac{1}{2}}|0\rangle=|T\rangle
%\end{equation}}}:
%\begin{equation}
%\begin{aligned}
%     &\cE(x_1,y_1)\cE(x_2,y_2)\\
%     \sim & \frac{1}{4(x_1-x_2)^2}+\frac{\psi_1(x_1,y_1)\psi_0(x_2,y_2)+\psi_0(x_1,y_1)\psi_1(x_2,y_2)}{4(x_1-x_2)}\\
%      \sim  &\frac{1}{4(x_1-x_2)^2}+\frac{(x_1-x_2)\partial_1\psi_1(x_2,y_2)\psi_0(x_2,y_2)+(x_1-x_2)\partial_1\psi_0(x_2,y_2)\psi_1(x_2,y_2)}{4(x_1-x_2)}\\
%     \sim & \frac{1}{4(x_1-x_2)^2}+\frac{1}{2}T(x_2,y_2)
%\end{aligned}
%\end{equation}
%}{}

We can write down the modes expansion of these four operators, %they are (we also write the one for $\epsilon$, which is more fundamental than the one for its derivatives):
\begin{equation}\label{modeMKT}
\begin{aligned}
  K(x,y)&=\sum_nK_n^{(0)}x^{-n-2}+\sum_nK_n^{(1)}x^{-n-3}y+\sum_nK_n^{(2)}x^{-n-4}y,\\
  T(x,y)&=\sum_nT_n^{(0)}x^{-n-2}+\sum_nT^{(1)}_nx^{-n-3}y,\\
  M(x)&=\sum_nM_n^{(0)}x^{-n-2},\\
  \partial_1\cE(x,y)&=\sum_n\cE^{(0)}_n(-n-1)x^{-n-2}+\sum_n\cE^{(1)}_n(-n-2)x^{-n-3}y,
\end{aligned}
\end{equation}
where
\begin{equation}
\begin{aligned}
  K^{(0)}_n&=-\sum_mA_{n-m}A_m(-m-\frac{1}{2}),\\
  K^{(1)}_n&=-\sum_mB_{n-m}A_m(m+\frac{1}{2})(n-m+\frac{1}{2})-\sum_mA_{n-m}B_m(m+\frac{1}{2})(m+\frac{3}{2}),\\
  K^{(2)}_n&=\sum_mB_{n-m}B_m(n-m+\frac{1}{2})(m+\frac{1}{2})(m+\frac{3}{2}),\\
  L_n\equiv T^{(0)}_n&=-\frac{1}{2}\sum_mA_{n-m}B_m(-m-\frac{1}{2})-\frac{1}{2}\sum_mB_{n-m}A_m(-m-\frac{1}{2}),\\
  T^{(1)}_n&=(-n-2)M^{(0)}_n\\
  M_n\equiv M^{(0)}_n&=-\frac{1}{2}\sum_mB_{n-m}B_m(-m-\frac{1}{2}),\\
  -J_n\equiv \cE^{(0)}_n&=-\frac{1}{2}\sum_mB_{n-m}A_m,\\
  \cE^{(1)}_n&=M^{(0)}_n.
\end{aligned}\label{nmodes}
\end{equation}
 The states corresponding to the operators in the stress tensor multiplet  are
\begin{equation}
\begin{aligned}
  |K\rangle&=-A_{-\frac{1}{2}}A_{-\frac{3}{2}}|0\rangle,\\
  |T\rangle&=-\frac{1}{2}(A_{-\frac{1}{2}}B_{-\frac{3}{2}}+B_{-\frac{1}{2}}A_{-\frac{3}{2}})|0\rangle,\\
  |M\rangle&=-\frac{1}{2}B_{-\frac{1}{2}}B_{-\frac{3}{2}}|0\rangle,\\
  |\partial_1\cE\rangle&=-\frac{1}{2}(-A_{-\frac{1}{2}}B_{-\frac{3}{2}}+B_{-\frac{1}{2}}A_{-\frac{3}{2}})|0\rangle,
\end{aligned}
\end{equation}
which satisfy the following relations under  the action of $L_0$, $M_0$ and $J_0$:
\begin{equation}
 L_0|K\rangle=2|K\rangle, \qquad L_0|T\rangle=2|T\rangle, \qquad L_0|\partial_1\cE\rangle=2|\partial_1\cE\rangle, \qquad L_0|M\rangle=2|M\rangle,
\end{equation}
\begin{equation}
 M_0|K\rangle=2|T\rangle-|\partial_1\cE\rangle, \qquad M_0|T\rangle=2|M\rangle, \qquad M_0|\partial_1\cE\rangle=|M\rangle, \qquad M_0|M\rangle=0.
\end{equation}
\begin{equation}
 J_0|K\rangle=-|K\rangle, \qquad J_0|T\rangle=0, \qquad J_0|\partial_1\cE\rangle=0, \qquad J_0|M\rangle=|M\rangle,
\end{equation}
Therefore, the stress tensor operators, together with $K$ and $\partial_1 \cE$, form a BMS triplet $|\mathbb{T}^3\rangle$ and a BMS singlet $|\mathbb{T}^1\rangle$, where
\begin{equation}
  |\mathbb{T}^3\rangle=(\frac{3}{2}|M\rangle, |T\rangle-\frac{1}{2}|\partial_1\cE\rangle, \frac{1}{2}|K\rangle)^\top, \qquad  |\mathbb{T}^1\rangle=|T\rangle-2|\partial_1\cE\rangle
\end{equation}
with conformal dimensions, boost charges and $u(1)$ charges
\begin{equation}
 \Delta_{\mathbb{T}^3}=\left(
 \begin{matrix}
   2 & 0 &0\\
   0 &2 &0 \\
   0 & 0 &2
 \end{matrix}\right),
  \qquad \xi_{\mathbb{T}^3} =\left(
 \begin{matrix}
   0 & 0 &0\\
   1 &0 &0 \\
   0 & 1 &0
 \end{matrix}\right), \qquad
 \varepsilon_{\mathbb{T}^3} =\left(
 \begin{matrix}
   1 & 0 &0\\
   0 &0 &0 \\
   0 & 0 &-1
 \end{matrix}\right).
\end{equation}
\begin{equation}
      \Delta_{\mathbb{T}^1}=2, \qquad \xi_{\mathbb{T}^1}=0, \qquad \varepsilon_{\mathbb{T}^1}=0.
\end{equation}
It is worth remarking that $J_0$ provides a $u(1)$ charge, under which the states $|K\rangle$, $|M\rangle$ carry the charge $1$ and $-1$, respectively, while $|T\rangle$ and $|\cE\rangle$ are neutral.  This perspective will be useful in the following discussion on BMS-Kac-Moody staggered modules.

%It can be easily checked whether these states are primary or quasi-primary states with respect to the Virasoro/BMS/BMS-Kac-Moody algebras respectively.
With respect to the Virasoro subalgebra,  $|K\rangle$, $|T\rangle$, $|M\rangle$ are Viraroso quasi-primaries ($|K\rangle$ and $|M\rangle$ are in fact  Viraroso primaries), as
\begin{equation}\label{Misquasi1}
 L_1|K\rangle=L_1|T\rangle=L_1|M\rangle=0, \qquad  L_2|K\rangle=L_2|M\rangle=0.
\end{equation}
Note that $|M\rangle$ is a global BMS descendant\footnote{$|\cE\rangle$ is a BMS primary, so of course is a Viraroso primary.} of $|\cE\rangle$ but not as a global Viraroso descendant, so it can be a Viraroso  quasi-primary. $|\partial_1\cE\rangle$ is the global Viraroso descendant of $|\cE\rangle$, so is not a Viraroso  quasi-primary ($L_1|\partial_1\cE\rangle=2|\cE\rangle$).
 With respect to the BMS algebra, $|T\rangle$ and  $|M\rangle$ are  both BMS quasi-primaries
\begin{equation}\label{Misquasi2}
 M_1|T\rangle=M_1|M\rangle=L_1|T\rangle=L_1|M\rangle=0,
\end{equation}
however, $|K\rangle$ is not a BMS quasi-primary
\begin{equation}\label{Knotquasi}
 M_1|K\rangle=-2|\cE\rangle.
\end{equation}
Besides, though  $|\partial_1\cE\rangle$ is not a BMS quasi-primary, it is annihilated by $M_1$
\begin{equation}
 M_1|\partial_1\cE\rangle=M_1L_{-1}|\cE\rangle=2M_0|\cE\rangle=0.
\end{equation}
Finally, with respect to the BMS-Kac-Moody algebra, we have :
\begin{equation}
    J_1|M\rangle=J_1|K\rangle=0,\quad J_1
    |L\rangle=-|\cE\rangle, \quad J_1|\partial_1\cE\rangle=0
\end{equation}
so only $|M\rangle$ is a  BMS-Kac-Moody quasi-primary states.

Now, we can describe the structure of a BMS-Kac-Moody  staggered  module. First of all, let us recall the general definition of a staggered  module. A staggered module $\mathcal{S}$ for an algebra $\mathcal{A}$ is an  $\mathcal{A}$-module for which we have a short exact sequence:
\begin{equation}\label{staggered}
    0\to \mathcal{H}^L\xrightarrow[]{\iota} \mathcal{S}\xrightarrow[]{\pi}\mathcal{H}^R\to 0
\end{equation}
where $\mathcal{H}^L$ and $\mathcal{H}^R$ are referred to as left modules and right modules respectively, both being  highest weight modules\footnote{More generally, they are called standard modules, see\cite{Creutzig:2013hma}.}. 
There is another central requirement: there must be an element $\mathcal{C}$ in the algebra   $\mathcal{A}$ which is not diagonalisable  on the module $\mathcal{S}$. In LCFT, $\mathcal{C}$ will always  be $L_0$.  A staggered module in fact gives a way to ``glue'' two highest-weight modules.  More generally, one could consider indecomposable   modules constructed from more than two highest weight modules.
%}{May be this part about the staggered module can be put in the appendix.}

For the staggered modules in the BMS free fermion, the underlying algebra $\mathcal{A}$ will be the BMS-Kac-Moody algebra and the element $\mathcal{C}$ will be $M_0$. As we mentioned above, $J_0$ can be viewed as a $u(1)$ charge number operator. It  is actually diagonalisable on the full BMS-Kac-Moody staggered module, which can be verified by the commutators involving $J_0$ (\eqref{LKJ}, \eqref{JTM}). Furthermore, for a state $|\cO\rangle$ of $J_0$ charge $\varepsilon$, acting $M_n$ (or $K_m$) on it would give rise to a state of $J_0$ charge $\varepsilon+1$(or $-1$).  Therefore we may assign a grade on the BMS-Kac-Moody staggered module according to their $u(1)$ charges. Because the BMS-Kac-Moody algebra involves only $L_n$, $M_n$ and $J_n$,  all  states with $-\varepsilon\leq \Lambda, \Lambda\in \mathbb{N}$  form a BMS-Kac-Moody module, denoted by $\mathcal{S}_\Lambda$. All 
$\mathcal{S}_\Lambda$ except one case are in fact BMS-Kac-Moody staggered modules as we will show in the following. The exception comes from the case $\mathcal{S}_{-\varepsilon_0}$ with $\varepsilon_0$ being the $u(1)$ charge of the BMS-Kac-Moody primary state in the full staggered module, which is a BMS-Kac-Moody highest-weight module.%(I still feel that this paragraph is not clear)}

As an illustration, we consider the staggered vacuum module $\mathcal{V}$ in the BMS free fermion.  In this vacuum module, we have the submodule  $\mathcal{S}_1$, by its definition it can be  generated by\footnote{The module generated by \eqref{S1} of course contains   null states, which need to be modded out to obtain the module $\mathcal{S}_1$. }:
\begin{equation}\label{S1}
\begin{aligned}
     \{L_{n_1}&L_{n_2}...L_{n_i}M_{m_1}M_{m_2}...M_{m_j}J_{k_1}J_{k_2}...J_{k_j}K_{-2}^a
    |0\rangle\}\\
    -1\leq n_1\leq n_2\leq...\leq n_i,\quad &-1\leq m_1\leq m_2\leq...\leq m_j,\quad 
    -1\leq k_1\leq k_2\leq...\leq k_j ,\quad a=0, 1
\end{aligned}
\end{equation}
$\mathcal{S}_1$ is actually a  staggered module.  We can see the staggered structure as follows. In \eqref{S1},  the set of
vectors with $a=0$ form a submodule $\Omega\equiv \mathcal{S}_0$ (with null states being modded out), which is  the BMS-Kac-Moody  highest-weight vacuum module. Then we have the following short exact sequence:
\begin{equation}
    0\to \Omega\xrightarrow[]{\iota} \mathcal{S}_1\xrightarrow[]{\pi} \mathcal{K}_1\to 0
\end{equation} 
where $\mathcal{K}_1\equiv\mathcal{S}_1/\Omega$ is the quotient module. This quotient module can be generated by the generators in \eqref{S1} with $a=1$, and it indeed gives a BMS-Kac-Moody module. Furthermore, it is in fact a BMS-Kac-Moody highest-weight module with primary states $|\tilde{K}\rangle$, which is the quotient class including $|K\rangle$. Note that in $\mathcal{S}_1$ or $\mathcal{V}$, $|K\rangle$  is not a BMS-Kac-Moody primary. However, in the quotient module $\mathcal{K}_1$, we have   $\Omega\sim 0$, so $|\tilde{K}\rangle$ is indeed a primary
\begin{equation}
    M_2|\tilde{K}\rangle=M_1|\tilde{K}\rangle=L_2|\tilde{K}\rangle=L_1|\tilde{K}\rangle=J_1|\tilde{K}\rangle=0
\end{equation}
and it has the following charges
\begin{equation}
    L_0|\tilde{K}\rangle=2|\tilde{K}\rangle, \quad M_0|\tilde{K}\rangle=0, \quad J_0
    |\tilde{K}\rangle=-|\tilde{K}\rangle.
\end{equation}  
Due to  $\Omega\sim 0$, $|\tilde{K}\rangle$ is a singlet primary under $M_0$.
Therefore we can identify the left and right modules as $\mathcal{H}^L=\Omega$ and $\mathcal{H}^R=\mathcal{K}_1$ to recognize the staggered module $\mathcal{S}_1$. Now, regarding $\mathcal{S}_1$ as $\mathcal{H}^L$, we can further enlarge it by including the operator $\hat{N}(KK)$ (then $a=0,1,2$ in \eqref{S1}) to obtain the BMS-Kac-Moody module $\mathcal{S}_2$.  The  quotient module $\mathcal{K}_2\equiv \mathcal{S}_2/\mathcal{S}_1$ turns out to be a highest-weight module, which can be regarded as $\mathcal{H}^R$, then $\mathcal{S}_2$ is  a staggered module as well.  Similarly,  we can  recognize the staggered module structure of $\mathcal{S}_3$, $\mathcal{S}_4$, $\mathcal{S}_5$ .... Finally, the BMS-Kac-Moody vacuum module $\mathcal{V}$ will be realized as a BMS-Kac-Moody staggered module $\mathcal{S}_\infty$. This construction is in fact  similar with the one in the BMS free scalar model. 

The construction of  staggered modules in the BMS free field theories is remarkably different from the one in LCFT. In the BMS cases,  staggered modules glue infinite number of  highest-weight modules, while in the LCFT case,  staggered modules always  glue two   highest-weight modules\footnote{In the LCFT literature, higher rank Jordan blocks have also appeared and been investigated, but most authors only  refer to the rank 2 cases as  staggered modules. Here we borrow the terminology ``staggered module'' even for   infinite rank Jordan blocks, following \cite{Hao:2021urq} for the BMS free scalar.}. This remarkable feature indicates that $K$ is in fact a generator of an enlarged algebra.
%}{The discussion here is similar with the one in the BMS free scalar model. Personally, the  staggered module description used here and in the BMS free scalar model are not very satisfying: they glue , while in LCFT, staggered module always  glue 2 (finite) highest weight modules. This infinity is more like that $K$ is a generator of an enlarged algebra. }

For  general BMS-Kac-Moody staggered modules, we can similarly construct them   as $\mathcal{S}_\infty$. The only difference is that the initial $\mathcal{H}^L$ will be a general  BMS-Kac-Moody  highest weight module $\mathcal{O}$, and we enlarge it by $\hat{N}(K\mathcal{O})$, $\hat{N}(KK\mathcal{O})$, etc. The crucial point for this to work is that we need to know the commutators of the modes of $K$ with all the elements of the BMS-Kac-Moody algebra. In fact, in the BMS free fermion, all the modes form a $W$-algebra and the  BMS-Kac-Moody  staggered  modules can in fact be  viewed as  highest-weight modules of this $W$-algebra. We will show this later.

It would be interesting to compare the symmetry and the stress tensor multiplet with the ones in the BMS free scalar model \cite{Hao:2021urq}% Now let us list differences comparing to the BMS free scalar.
\begin{itemize}
    \item Firstly, let us compare the symmetry.   The BMS free scalar model also has a BMS-Kac-Moody algebra  as its underlying symmetry. The difference is that in that case, the Kac-Moody level vanishes so the $u(1)$ current does not enlarge the BMS highest weight modules. However, in our case the $u(1)$ Kac-Moody level does not vanish so the $u(1)$ current  indeed enlarge a BMS highest weight module to a BMS-Kac-Moody highest weight module (then it is further enlarged to a staggered module by including $K$). Besides, in the BMS free scalar, the $u(1)$ current $j$ comes from an internal symmetry: the invariance of the action under  $y$ independent shifts of the fundamental scalar field. In our case, the $u(1)$  current comes from the symmetry $\mathcal{C}_f$ in \eqref{cf1} and \eqref{Cf} which involving space-time transformations. 
    \item The symmetry algebras of both theories  contain only one Kac-Moody current, so  both can not be an algebra which comes from a contraction of  two (chiral and anti-chiral) relativistic Kac-Moody algebras. However, while the symmetry algebra found in \cite{Hao:2021urq} is in fact a subalgebra of a bigger algebra which can be obtained from a  contraction\footnote{This bigger algebra is generated by $O_0$ and $O_1$ in \cite{Hao:2021urq}. In fact, $T$ and $M$ can be constructed as normal ordered products of $O_0$ and $O_1$ (a Sugawara like construction). It is shown in \cite{Hao:2021urq} that this bigger algebra can be obtained by contracting two $u(1)$ Kac-Moody algebras. Note that $O_1$ is not a symmetry current, thus the symmetry algebra of the BMS free scalar is only a subalgebra of this bigger algebra.    }, our BMS-Kac-Moody algebra can not be such a subalgebra, as here  $[M_n,J_m]\varpropto M_{n+m}$(\eqref{JTM}).%, as we mentioned in the paragraph below \eqref{ClaBMSK}.}
    \item It is interesting to notice that in the BMS free scalar model,  the normal-ordered product of the $u(1)$ current $j$ gives $M$,
     \begin{equation}
         \hat{N}(jj)\sim M.
      \end{equation}
    While in our case,  the normal ordered product of the current $\cE$ gives $T$:
     \begin{equation}
         \hat{N}(\cE\cE)\sim T
       \end{equation}
    Remarkably,  one can also construct $M$ in terms of the current $\mathcal{E}$: not by the  Sugawara construction, but by a derivative $M=\partial_0\mathcal{E}$. This kind of construction may not apply to general BMS field theories but it really happens in the BMS free fermion case.\footnote{In fact, a similar relation with $M=\partial_0\mathcal{E}$ also appear in the BMS free scalar, see \eqref{scalarJM}. } At first sight, it seems strange that the stress tensor $M$ is a derivative of another current $\cE$, which means that $M$ is a global descendent of $\cE$. In the CFT case, such an operator cannot be a quasi-Virasoro primary. Interestingly, one can easily verify that $M$  is indeed a BMS quasi-primary, and we had stressed this fact in \eqref{Misquasi1}, \eqref{Misquasi2} and the paragraph between them.  Logically, the  relation $\partial_0\cE=M$  precisely reflects the fact that there exist a very special enlarged BMS algebra, which enlarge the BMS algebra by the current $\cE$. This algebra is just the new type of BMS-Kac-Moody algebra we found in this paper.
    \item In the case of the BMS free scalar, $T, M$ and $K$ form a triplet. In our case,  roughly speaking, $K$, $T$, $M$ still form a triplet. However, due to the existence of the fourth operator $\partial_1\cE$, the triplet gets modified slightly and there appears  another new singlet $|\mathbb{T}^1\rangle$.
    \item While the operators in the BMS free scalar model can not be simply organized into BMS primaries and their descendants, they can be organized into BMS quasi-primaries and their global descendants. More precisely,  BMS highest weight modules  must be enlarged to   BMS  staggered modules. For example,  the operator $K$ is a BMS quasi-primaries but not as a descendant of any BMS primaries. The case in the BMS free fermion seems more novel: the operator $K$ is neither BMS quasi-primaries, nor a descendant of any BMS primary. In fact, this is because the enlarged module is not a BMS staggered module but a BMS-Kac-Moody staggered module.  In the case of the BMS free scalar, because the Kac-Moody level vanish,  the expected BMS-Kac-Moody staggered modules reduce to  BMS  staggered modules.
    \item In the BMS free fermions case, there are operators which belong to two different BMS modules, and  such a phenomenon does not appear in the BMS free scalar model.
    For example, the stress tensor $|M\rangle=\lim\limits_{x\to 0}M(x)|0\rangle=M_{-2}|0\rangle$ is a descendant of the vacuum $|0\rangle$, but at the same time it is a (global)  descendant of the primary state $|\cE\rangle$:
   \begin{equation}
    |M\rangle=M_{-1}|\cE\rangle.
   \end{equation}
   This fact reflects that the vacuum state $|0\rangle$ and the BMS primary state $|\cE\rangle$ indeed belong to a same enlarged BMS module: the BMS-Kac-Moody (staggered) vacuum module.
\end{itemize}

\subsubsection{Operator product expansion }\label{operatorexp}
Now we calculate some OPEs of the currents in the BMS free fermion theory. Using the Wick theorem  we can easily write down all the OPEs among various operators.
For $T$ and $M$, they are:
\begin{equation}
  \begin{aligned}
   & T(x_1,y_1)T(x_2,y_2)\sim \frac{1}{2(x_1-x_2)^4}+\frac{2T(x_2,y_2)}{(x_1-x_2)^2}-\frac{4(y_1-y_2)M(x_2,y_2)}{(x_1-x_2)^3}+ \frac{\partial_xT(x_2,y_2)}{x_1-x_2}-\frac{(y_1-y_2)\partial_yT(x_2,y_2)}{(x_1-x_2)^2},  \\
   & T(x_1,y_1)M(x_2,y_2)\sim \frac{2M(x_2,y_2)}{(x_1-x_2)^2}+\frac{\partial_xM(x_2,y_2)}{x_1-x_2},\\
   &M(x_1,y_1)M(x_2,y_2)\sim 0.
  \end{aligned}
\end{equation}
%\ZF{}{Agree with the revision here.}
Translating into the modes, we can check that the modes $L_n$ and $M_n$  satisfy the BMS algebra
\begin{equation}\label{TM}
\begin{aligned}
  \left[L_n,L_m\right]&=(n-m)L_{n+m}+\frac{c_L}{12}n(n^2-1),\\
  [L_n,M_m]&=(n-m)M_{n+m}+\frac{c_M}{12}n(n^2-1),\\
  [M_n,M_m]&=0.
\end{aligned}
\end{equation}
with  $c_L=1$ and $c_M=0$ being the central charges of the BMS free fermion theory. To obtain the Kac-Moody part of the BMS-Kac-Moody algebra, we need to write down the OPEs among  $T$, $M$ and $\cE$, they are:
\begin{equation}
\begin{aligned}
    T(x_1,y_1)\cE(x_2,y_2)&\sim \frac{\cE}{(x_1-x_2)^2}+\frac{\partial_1\cE}{x_1-x_2}-\frac{(y_1-y_2)\partial_0\cE}{(x_1-x_2)^2}\\
    M(x_1,y_1)\cE(x_2,y_2)&\sim \frac{\partial_0\cE}{x_1-x_2}\\
    \cE(x_1,y_1)\cE(x_2,y_2)&\sim \frac{1}{4(x_1-x_2)^2}
\end{aligned}
\end{equation}
Translating them into the modes, we have the commutation relations
\begin{equation}\label{JTM}
\begin{aligned}
    \left[L_n, J_m\right]&=-mJ_{n+m}\\
    [M_n, J_m]&=-M_{n+m}\\
    [J_n,J_m]&=\frac{1}{4}n\delta_{n+m,0}.
\end{aligned}
\end{equation}
These commutators, together with \eqref{TM}, form a BMS-Kac-Moody algebra with $c_L=1$, $c_M=0$
and the Kac-Moody level $k=\frac{1}{4}$. Note that because this is a $u(1)$ Kac-Moody, the  value of the Kac-Moody level is in fact a convention of choice: it can be any other value, except zero, by rescaling the current $J$ (at the same time change the commutator of $M_n$ and $J_m$, of course). It is worth emphasizing  that the Kac-Moody level here does not vanish.

It is easy to work out the OPE of the fundamental fermions to confirm that they form a doublet of  BMS-Kac-Moody primaries. The OPEs are of the following forms
\begin{equation}
\begin{aligned}
    T(x_1,y_1)\psi_0(x_2, y_2)&\sim \frac{\frac{1}{2}\psi_0}{(x_1-x_2)^2}+\frac{\partial_1\psi_0}{x_1-x_2}\\
     T(x_1,y_1)\psi_1(x_2, y_2)&\sim \frac{\frac{1}{2}\psi_1}{(x_1-x_2)^2}+\frac{\partial_1\psi_1}{x_1-x_2}-\frac{(y_1-y_2)\psi_0}{(x_1-x_2)^3}-\frac{(y_1-y_2)\partial_0\psi_1}{(x_1-x_2)^2}\\
      M(x_1,y_1)\psi_0(x_2, y_2)&\sim 0\\
       M(x_1,y_1)\psi_1(x_2, y_2)&\sim \frac{\frac{1}{2}\psi_0}{(x_1-x_2)^2}+\frac{\partial_0\psi_1}{x_1-x_2}
\end{aligned}
\end{equation}
As a consistent check of the symmetry $\mathcal{C}_f$,  we can calculate the OPE of $J=J^0_{\mathcal{D}'}=-\cE^{(0)}$ and $\psi_0$ or $\psi_1$:
       \begin{equation}\label{ope0}
        J\psi_0\sim \frac{\frac{1}{2}\psi_0}{x_1-x_2}
     \end{equation}
     \begin{equation}\label{ope1}
        J\psi_1\sim \frac{-\frac{1}{2}\psi_1}{x_1-x_2}
        +y_2\frac{\partial_1\psi_0}{x_1-x_2}+y_2\frac{\frac{1}{2}\psi_0}{(x_1-x_2)^2}
     \end{equation}
     From the leading terms in \eqref{ope0} and \eqref{ope1} one can directly read that the charge of $\psi_0$ and $\psi_1$ under $\mathcal{D}'$ are $\frac{1}{2}$ and $-\frac{1}{2}$, respectively. More precisely, we can calculate the infinitesimal variation of the fundamental fields from the above OPEs,
     \begin{equation}\label{omega}
         \delta_\omega \psi_i(x_2,y_2)=\frac{1}{2\pi i}\oint dx_1\omega(x_1)[-J(x_1)\psi_i(x_2,y_2)]
     \end{equation}
     with $f(x)=1+\omega (x)$. In fact, substituting \eqref{ope0} into  \eqref{omega}, we find
     \begin{equation}
          \delta_\omega \psi_0=\frac{1}{2}\omega\psi_0
     \end{equation}
     which is just the infinitesimal form of \eqref{Cf}  for $\psi_0$. Note that there are no derivative terms with respect to both $x$ and $y$ because
     \begin{equation}
         \delta x=x'-x=0, \qquad \partial_0\psi_0=0.
     \end{equation}
     Similarly,  substituting \eqref{ope1} into  \eqref{omega}, we find
     \begin{equation}
         \delta_\omega \psi_1=-\frac{1}{2}\omega\psi_1+y\omega \partial_0\psi_1+\frac{1}{2}y\omega'\psi_0,
     \end{equation}
      which is just the infinitesimal form of \eqref{Cf} for $\psi_1$.

The OPEs involving the operator $K$ are  more complicated, and they present a novel feature: the OPE can not be organized by BMS(-Kac-Moody) quasi-primaries. Firstly, note that $K$ itself is not a BMS quasi-primary. As $K$ is a Virasoro primary, one may try to use only the Virasoro subalgebra to organize the OPE. However, there exist operators in OPEs which are not Virasoro quasi-primaries themselves and are also not the derivatives of any other quasi-primaries appearing in the OPE. For example, let us look at the OPE of $M(x)K(x,y)$, which is of the form
\begin{equation}\label{MK}
   M(x_1,y_1)K(x_2,y_2)
  \sim\frac{1/2}{(x_1-x_2)^4}+\frac{-2\cE(x_2,y_2)}{(x_1-x_2)^3}+\frac{2T(x_2,y_2)-\partial_1\cE(x_2,y_2)}{(x_1-x_2)^2}+\frac{\partial_yK(x_2,y_2)}{x_1-x_2}.
\end{equation}
On the right-hand side, we  find an operator (and its corresponding state)
\begin{equation}
 K^{(1)}(x,y)\equiv  \partial_yK(x,y)\leftrightarrow M_{-1}|K\rangle,
\end{equation}
 which is neither a Virasoro (quasi-)primary as
 \begin{equation}
     L_1 M_{-1}|K\rangle=2M_0|K\rangle=4|T\rangle-2|\partial_1\cE\rangle\neq 0,
 \end{equation}
 nor the derivatives of  any Virasoro quasi-primaries appearing in the OPE. However, $K^{(1)}(x,y)$ can in fact be expressed as a sum of the derivatives of  Virasoro quasi-primaries.  It turns out that
\begin{equation}\label{K1}
    K^{(1)}(x,y)=\partial_1 T-\frac{1}{3}\partial_1^2\cE-\frac{16}{3}\hat{N}(\cE\cE\cE)
\end{equation}
where  $T$, $\cE$ and $\hat{N}(\cE\cE\cE)$ are all Virasoro quasi-primaries. Note that while $T$ and $\cE$ are BMS quasi-primaries, $\hat{N}(\cE\cE\cE)$ is not. So we conclude that while the OPE can not be organized by the  BMS quasi-primaries, it can be organized by the Virasoro quasi-primaries. %(That is, operator products of any Virasoro quasi-primaries can be expanded  in terms of  Virasoro quasi-primaries and their derivatives). }{}

 To find the corresponding algebra, we must translate \eqref{MK} into  the commutators of the modes. Firstly, we have
 \begin{equation}
 \begin{aligned}
     \frac{1}{2\pi i}&\oint dx x^{n+1}M(x)=M_n,\\
     \frac{1}{2\pi i}&\oint dx x^{m+1}K(x,y)=K^{(0)}_m+yK^{(1)}_{m-1}+y^2K^{(2)}_{m-2}.
 \end{aligned}    
 \end{equation}
 Then using the OPE \eqref{MK}, we find
\begin{equation}
\begin{aligned}
    &[M_n, K^{(0)}_m+yK^{(1)}_{m-1}+y^2K^{(2)}_{m-2}]\\
     =&\oint dx_1 x_1^{n+1}\oint dx_2 x_2^{m+1}\left[\frac{1/2}{(x_1-x_2)^4}+
     \frac{-2\cE(x_2,y_2)}{(x_1-x_2)^3}+\frac{2T(x_2,y_2)-\partial_1\cE(x_2,y_2)}{(x_1-x_2)^2}+\frac{\partial_yK(x_2,y_2)}{x_1-x_2}\right].
\end{aligned}
\end{equation}  
Comparing the terms with the same $y$-dependence, $y^0, y^1$ or $y^2$, we can read the commutators $[M_n, K^{(0)}_m], [M_n, K^{(1)}_m]$ and $[M_n, K^{(2)}_m]$, respectively. Most importantly, we have
\begin{equation}\label{Mkmodes}
    [M_n,K^{(0)}_m]=2(n+1)L_{m+n}-(n+1)(m+1)J_{n+m}+K^{(1)}_{n+m}+\frac{1}{12}n(n^2-1)\delta_{n+m,0}.
\end{equation}
Using \eqref{K1}, we can rewrite the modes $K^{(1)}_{n+m}$ as
\begin{equation}
    K^{(1)}_{n}=(-n-2)L_n+\frac{1}{3}(-n-1)(-n-2)J_n+\frac{16}{3}\hat{N}(JJJ)_n, 
\end{equation}
Substituting it into \eqref{Mkmodes}, we find the commutator
\begin{equation}\label{MKcomplete}
    [M_n,K_m]=(n-m)L_{n+m}+\frac{1}{3}(n^2+m^2-nm-1)J_{n+m}+\frac{16}{3}\hat{N}(JJJ)_{n+m}+
    \frac{1}{12}n(n^2-1)\delta_{n+m,0},
\end{equation}
where we denote $K^{(0)}_m$ as $K_n$. 
Similarly, one can work out $[M_n, K^{(1)}_m], [M_n, K^{(2)}_m]$. However, since $K^{(1)}_m$ and $ K^{(2)}_m$ are not  Virasoro quasi-primary modes, we do not need these commutators  to define the  algebra. In fact, we only need the commutators among those modes which are not only Virasoro quasi-primary modes, but also the generators. It turns out that these generators include $M_n$, $K_n\equiv K^{(0)}_n$ and $J_n$. Note that $L_n$ is not a generator because $L_n=2\hat{N}(JJ)_n$, even though  we will also write down the commutators involve $L_n$. The commutators among $L_n$, $M_n$ and $J_n$ were already worked out in \eqref{TM} and \eqref{JTM}, known as the BMS-Kac-Moody algebra. The most non-trivial commutator involving $K_n$ was worked out in \eqref{MKcomplete}. The remaining ones are $[L_n, K_m]$, $[K_n, J_m]$ and $[K_n, K_m]$, which can be easily read off from the OPE of $TK$, $KJ$ and $KK$, respectively:
\begin{equation}\label{LKJ}
\begin{aligned}
    \left[L_n, K_m\right]&=(n-m)K_{n+m}\\
    [K_n, J_m]&=K_{n+m}\\
    [K_n, K_m]&=0
\end{aligned}
\end{equation}
Note that commutators involving $K_n$ are similar with the ones involving $M_n$. A minor difference is an opposite sign appearing in the right hand sides of $[K_n, J_m]$  and $[M_n, J_m]$. This sign difference is not a convention of choice, and reflects the structure of the algebra. In \eqref{compareMK}, it may seems that $M$ and $K$ are defined with a different normalization so they are not   treated on a equal footing. However, this is  not the case because under the rescaling: $M_n\to \alpha M_n$, $K_n\to \alpha^{-1}K_n$,  all  commutators do not change.

It can be checked that   \eqref{TM},  \eqref{JTM}, \eqref{MKcomplete} and \eqref{LKJ} indeed define a non-linear algebra,  which we denote as $\mathfrak{S}$. It is a $W$-algebra of type $W(2,2,1)$, generated by three Virasoro primaries $M_n$, $K_n$ and $J_n$. In fact,  the BMS-Kac-Moody staggered modules appearing in the BMS free
 fermion can be viewed as  the highest-weight modules of $\mathfrak{S}$. To see this, note that a highest-weight (singlet) module of the W-algebra $\mathfrak{S}$ is represented by a $\mathfrak{S}$-primary state defined as
 \begin{equation}
     M_n|\mathcal{O}\rangle=K_n|\mathcal{O}\rangle=J_n|\mathcal{O}\rangle=L_n|\mathcal{O}\rangle=0, \qquad n>0,
 \end{equation}
 and is labeled by the eigenvalues of $M_0,L_0,J_0$ and $K_0$
 \begin{equation}
     M_0|\mathcal{O}\rangle=\xi|\mathcal{O}\rangle, \quad L_0|\mathcal{O}\rangle=\Delta|\mathcal{O}\rangle,\quad J_0|\mathcal{O}\rangle=\varepsilon|\mathcal{O}\rangle,\quad
     K_0|\mathcal{O}\rangle=\kappa|\mathcal{O}\rangle.
 \end{equation}
 It is easy to see that the  BMS-Kac-Moody staggered vacuum module in the BMS free fermion is the highest-weight vacuum module of $\mathfrak{S}$ with
 \begin{equation}
     \xi=\Delta=\varepsilon=\kappa=0. 
 \end{equation}
 
Let us look at the zero modes of the $W$-algebra $\mathfrak{S}$ more carefully. The zero modes have the following commutation relations
 \begin{equation}\label{zero1}
     [L_0, J_0]=[L_0, M_0]=[L_0, K_0]=0
 \end{equation}
  \begin{equation}\label{zero2}
     [J_0, M_0]=M_0, \quad [J_0, K_0]=-K_0, \quad [M_0, K_0]=-\frac{1}{3}J_0+\frac{16}{3}J_0J_0J_0.
 \end{equation}
 These zero modes do not commute mutually, which indicates that there could exist multiplet structure (degenerate ground states). For the  singlet case, there is only one primary state   $|\mathcal{O}\rangle$  with dimension $\Delta_{\mathcal{O}}$, $M_0$ charge $\xi_{\mathcal{O}}$, $K_0$ charge $\kappa_{\mathcal{O}}$ and $u(1)$ charge $\varepsilon_{\mathcal{O}}$, and we find the following relations
 \begin{equation}
     0=(\varepsilon_{\mathcal{O}}\xi_{\mathcal{O}}-\xi_{\mathcal{O}}\varepsilon_{\mathcal{O}})
     |\mathcal{O}\rangle=[J_0, M_0]|\mathcal{O}\rangle=M_0|\mathcal{O}\rangle=\xi_{\mathcal{O}}
     |\mathcal{O}\rangle,\nn
 \end{equation}
  \begin{equation}\nn
     0=(\varepsilon_{\mathcal{O}}\kappa_{\mathcal{O}}-\kappa_{\mathcal{O}}\varepsilon_{\mathcal{O}})
     |\mathcal{O}\rangle=[J_0, K_0]|\mathcal{O}\rangle=-K_0|\mathcal{O}\rangle=\kappa_{\mathcal{O}}
     |\mathcal{O}\rangle,
 \end{equation}
 \begin{equation}\label{M0K0}
    0=(\xi_{\mathcal{O}}\kappa_{\mathcal{O}}-\kappa_{\mathcal{O}}\xi_{\mathcal{O}})|\mathcal{O}\rangle= [M_0, K_0]|\mathcal{O}\rangle=(-\frac{1}{3}J_0+\frac{16}{3}J_0J_0J_0)|\mathcal{O}\rangle=(-\frac{1}{3}\varepsilon_{\mathcal{O}}+\frac{16}{3}
    \varepsilon_{\mathcal{O}}^3)|\mathcal{O}\rangle,
 \end{equation}
 which lead to $\xi_{\mathcal{O}}=\kappa_{\mathcal{O}}=0$ and $\varepsilon_{\mathcal{O}}=0,\pm\frac{1}{4}$. Besides the BMS-Kac-Moody vacuum module which has $\varepsilon_{\mathcal{O}}=0$, there are two other cases with $\varepsilon_{\mathcal{O}}=\pm\frac{1}{4}$. In fact, they correspond exactly to the two
 twist operators $\mu$ and $\sigma$ in the Ramond sector, which will be discussed in the next section. For Viraroso-Kac-Moody primaries, we have $\Delta=2\varepsilon^2$ due to the relation\footnote{It is clear that $T$ and $\cE$ do not satisfy $\Delta=2\varepsilon^2$ because they are not Virasoro-Kac-Moody primaries. The operators $K$, $M$, $\psi_0$ and $\psi_1$ satisfy this relation because they are all Virasoro-Kac-Moody primaries.} $T=2\hat{N}(\cE\cE)$. Thus these two twist operators both have dimension $\Delta=2\times (\pm\frac{1}{4})^2=\frac{1}{8}$.
 
In the NS sector of the BMS free fermion, there is  another BMS-Kac-Moody staggered module $\mathcal{F}$ including the fundamental fermions, which form a  BMS-Kac-Moody primaries doublet: $(\psi_0, \psi_1)^{\mathrm{T}}$. We has discussed general  BMS-Kac-Moody staggered modules containing one BMS-Kac-Moody primary singlet state. The structure of staggered modules containing   BMS-Kac-Moody primary multiplets is similar.   In fact, these modules  can be understood better by realizing it as a highest-weight module of the $W$-algebra $\mathfrak{S}$. More precisely,  $\psi_0$ and $\psi_1$ form a $\mathfrak{S}$-primary doublet, which has the following representations  of the algebra of the zero modes \eqref{zero1}, \eqref{zero2}, 
 \begin{equation}\label{doublet0}
  \Delta=\left(\begin{matrix}
   \frac{1}{2} &0  \\
   0& \frac{1}{2}\end{matrix}\right), \qquad
   \xi=\left(\begin{matrix}
   0 &0  \\
   \frac{1}{2}& 0\end{matrix}\right), \qquad
    \kappa=\left(\begin{matrix}
   0 &1  \\
   0& 0\end{matrix}\right), \qquad
   \varepsilon=\left(\begin{matrix}
   \frac{1}{2} &0  \\
   0& -\frac{1}{2}\end{matrix}\right).
\end{equation}
Thus the BMS-Kac-Moody staggered module $\mathcal{F}$ is a  highest-weight module of the algebra $\mathfrak{S}$, as a   $\mathfrak{S}$-primary doublet $(\psi_0, \psi_1)^{\mathrm{T}}$.

 This  $W$-algebra $\mathfrak{S}$ is in fact a quantum version of the  conformal BMS algebra (CBMS) studied in \cite{Fuentealba:2020zkf}.  In \cite{Fuentealba:2020zkf}, it was found that the classical  conformal BMS algebra can be defined for generic central charge $c$.  For generic $c$, $T$ is also a generator of the $W$-algebra so the classical  conformal BMS algebra is of type $W(2,2,2,1)$.  Quantum conformal BMS algebra can be obtained by the quantum Drinfeld-Sokolov reduction based on a non-principal $sl_2$ embedding (see \cite{Fuentealba:2020zkf, 2022CMaPh.390...33F} or Appendix  \ref{boo}) so the central charge $c\equiv c_L$ can be generic.  In appendix \ref{boo}, we  show that when this algebra has a BMS-Kac-Moody subalgebra, the central charge has to be $c=1$ and the Kac-Moody level must be $k=\frac{1}{4}$. In this case, the current $T$ is decoupled, and we reproduce the W-algebra $\mathfrak{S}$ of type $W(2,2,1)$. The detailed discussions on this issue can be found in Appendix  \ref{boo}. Under this identification, the currents $\cE$ and $K$ could be interpreted as ``super-dilation'' and `` super-special conformal transformation'' operators in the bulk \cite{Fuentealba:2020zkf}.

 %It involves the modes $K^{(1)}_n$ and $K^{(2)}_n$, which do not appear in the one organized by the modes of Virasoro quasi-primaries including only \footnote{So we simply denote $K^{(0)}_n$ as $K_n$ in the bottom up construction of the W algebra in section \ref{boo} } and  \footnote{Generally, there will be some currents involving derivatives, but in the case at hand, no such terms appear.} so it is a W algebra of type W(2,2,2,1). $K^{(1)}_n$ and $K^{(2)}_n$ can be represented by these generating modes, so this novel algebra of modes is just a `bad' rewriting of the underlying W(2,2,2,1).

If we focus on the  modes from the beginning, then \eqref{complexfermodes} is just the algebra of a  complex fermion.  The W-algebra $\mathfrak{S}$  is the maximal bosonic subalgebra of this complex fermion algebra.
This $W$-algebra in fact encodes the information of the vacuum module, just like the usual 2D CFT case when the symmetry get enhanced. However, because the vacuum module here is a BMS-Kac-Moody staggered module,
the BMS free fermion actually gives a novel current realization of  the $W$-algebra. Firstly,  while   $T$, $M$ and $\cE$ indeed encode the BMS-Kac-Moody symmetry,  the operator $K$, which lie in the vacuum module and is a generator  of this W-algebra, does not encode any  symmetry transformation for the BMS free fermion\footnote{We currently do not find the possible  symmetry transformation corresponding to the operator $K$. This is why we  need to use  BMS-Kac-Moody staggered modules to organized the Hilbert space. We believe that $K$ can not be realized as a symmetry current in the BMS free fermion. This is very likely also the case for the BMS free scalar model \cite{Hao:2021urq}.}. Secondly,   the OPEs of the operators in the BMS free fermion can not be organized by  BMS quasi-primaries. Nevertheless, they can be organized by  Virasoro quasi-primaries as we have shown above.

Now, we want to show the non-unitarity of the BMS free fermion. Firstly, from the Hermitian condition  \eqref{hermitian}, we have
\begin{equation}\label{Her}
    L_n^\dagger=L_{-n}, \quad M_n^\dagger=M_{-n}, \quad K_n^\dagger=K_{-n},\quad J_n^\dagger=-J_{-n},
\end{equation}
so when $\Delta=1$, there is only one state $|\cE\rangle$, whose norm is negative,
\begin{equation}
    \langle\cE|\cE\rangle=\langle 0|J^{\dagger}_{-1}J_{-1}|0\rangle=-\frac{1}{4}.
\end{equation}
Similarly, when $\Delta=2$, the Gram matrix of the four independent states
\begin{equation}
    \left(\begin{matrix}
        \langle M| M\rangle & \langle M|K\rangle &\langle M|L\rangle & \langle M|\partial_1\cE\rangle\\
       \langle K| M\rangle  & \langle K| K\rangle  &\langle K| L\rangle  & \langle K| \partial_1\cE\rangle \\
        \langle L| M\rangle  & \langle L| K\rangle  & \langle L| L\rangle  & \langle L| \partial_1\cE\rangle \\
       \langle \partial_1\cE| M\rangle  & \langle \partial_1\cE| K\rangle & \langle \partial_1\cE| L\rangle &  \langle \partial_1\cE| \partial_1\cE\rangle \\
    \end{matrix}
    \right)
    = \left(\begin{matrix}
        0 & \frac{1}{2} & 0 & 0\\
       \frac{1}{2}  & 0 & 0  & 0 \\
        0 & 0  & \frac{1}{2}  & 0 \\
       0  & 0 & 0 &  -\frac{1}{2} \\
    \end{matrix}
    \right)
\end{equation}
is not positive definite. As a result, there are also the states with negative norms, for example, $|M\rangle-|K\rangle$ and $|\partial_1\cE\rangle$. 
The above states  lie in the vacuum module $\mathcal{V}$. One can check that there are the states with  negative norms in the module $\mathcal{F}$ as well.
%}{This part perhaps can be put in other places, currently I put it here.}

Finally, we would like to comment on the Hermitian condition \eqref{Her}. We have shown that the generators $J_n$, $M_n$, $K_n$ and $L_n$ form a non-linear $W$-algebra $\mathfrak{S}$.  Usually, the $W$-algebra appearing in a 2d unitary CFT can be equipped with an Hermitian condition, which
\begin{itemize}
    \item gives rise to a positive definite Gram matrix,
    \item is compatible with the structure of the $W$-algebra.
\end{itemize}
 In the case at hand,   such an  Hermitian condition\footnote{This Hermitian condition can  be obtained formally by imposing $A_n^\dagger=B_{-n}, B_n^\dagger=A_{-n}$ in  \eqref{nmodes}.} actually exists for  $\mathfrak{S}$:
\begin{equation}\label{otherH}
    M_n^\dagger=K_{-n}, \quad K_n^\dagger=M_{-n}, \quad L_n^\dagger=L_{-n}, \quad J_n^\dagger=J_{-n}.
\end{equation}
One can check the above two requirements are indeed satisfied by \eqref{otherH}. On the other hand, the Hermitian condition \eqref{Her} is also compatible with $\mathfrak{S}$. In other words, the $W$-algebra $\mathfrak{S}$ in fact admits two different Hermitian conditions, and the BMS free fermion  presents the non-unitary one.

\section{Ramond sector and the BMS Ising model}\label{RamondIsing}

 Recall that the usual two dimensional Ising model can be represented by the Majorana
fermion as follows\footnote{The standard  normalization $\langle\epsilon|\epsilon\rangle=1$ requires a factor $i$.},
\begin{equation}
  \mathbf{1}=\mathbf{1}_{\text{free fermion}}, \quad \epsilon=i\hat{N}(\psi\bar{\psi}), \quad \sigma=\text{twist operator in the Ramond sector.}
\end{equation}
We will show in this section that there is a similar BMS free fermion realization of the BMS Ising model.
To find such a realization  we essentially need to discuss the Ramond (R) sector to find the `spin operator' $\sigma$.

\subsection{Twist operators}
In the R-sector, the modes number $n\in\mathbb{Z}$ so the fermions satisfy
\begin{equation}
 \psi_0(e^{2\pi i}x)=-\psi_0(x), \qquad \psi_1(e^{2\pi i}x,y)=-\psi_1(x,y).
\end{equation}
Firstly, we need to find the degenerate ground states in the Ramond sector. The R-sector ground states are created by the twist operators.  For example, in the R-sector of the free Majorana fermion, one find two twist field\footnote{We will use the same notation $\sigma$ and $\mu$ for the twist operators in the BMS case.} $\sigma$ and $\mu$ from a two-dimensional representation of the Clifford algebra coming from the zero modes. They transform into each other when fusing with the two fundamental fermions $\psi$ and $\bar{\psi}$. In the BMS case, we also need to study the algebra of the zero modes. Recalling that
\begin{equation}
  \{A_n,A_m\}=0, \qquad \{B_n,B_m\}=0, \qquad \{A_n,B_m\}=\delta_{n+m,0},
\end{equation}
we combine them as
\begin{equation}
 C_n=\frac{1}{\sqrt{2}}(A_n+B_n), \qquad D_n=\frac{i}{\sqrt{2}}(A_n-B_n),
\end{equation}
and then find  that they satisfy the same algebra as the modes of free fermions $\psi_n$ and $\bar{\psi}_m$:
\begin{equation}
  \{C_n,C_m\}=\delta_{n+m,0}, \qquad \{D_n,D_m\}=\delta_{n+m,0}, \qquad \{C_n,D_m\}=0.
\end{equation}
For the zero modes $A_0,B_0$, they are  transformed  into  $C_0,D_0$, which obey
the same Clifford algebra just as $\psi_0$ and $\bar{\psi}_0$,
\begin{equation}
 C_0^2=\frac{1}{2}, \qquad D_0^2=\frac{1}{2}, \qquad \{C_0, D_0\}=0. 
\end{equation}
The zero modes as well as the fermionic number operator $(-1)^F$, which is used to distinguish the degenerate ground states,  can be realized in terms of the Pauli matrices:
\begin{equation}
\begin{aligned}
  &C_0=\frac{\sigma_x+\sigma_y}{2}(-1)^{\sum_{n>0}C_{-n}C_n+D_{-n}D_n},\\
  &D_0=\frac{\sigma_x-\sigma_y}{2}(-1)^{\sum_{n>0}C_{-n}C_n+D_{-n}D_n},\\
  &(-1)^F=\sigma_z(-1)^{\sum_{n>0}C_{-n}C_n+D_{-n}D_n}.
\end{aligned}
\end{equation}
Now we have two twist fields $\sigma$ and $\mu$, creating two R-sector ground states
\be
|0\rangle_{\sigma}\equiv\sigma(0,0)|0\rangle=|\sigma\rangle,\hs{3ex} |0\rangle_{\mu}\equiv\mu(0,0)|0\rangle=|\mu\rangle.
\ee 
These two ground states are transformed into each other by the zero modes, 
\begin{equation}
\begin{aligned}
 C_0|\sigma\rangle &=\frac{1-i}{2}|\mu\rangle, \qquad C_0|\mu\rangle=\frac{1+i}{2}|\sigma\rangle,\\
 D_0|\sigma\rangle &=\frac{1+i}{2}|\mu\rangle, \qquad D_0|\mu\rangle=\frac{1-i}{2}|\sigma\rangle.
\end{aligned}
\end{equation}
In terms of $A_0$  and $B_0$, we have
\begin{equation}
\begin{aligned}
 A_0|\sigma\rangle &=\frac{1-i}{\sqrt{2}}|\mu\rangle, \qquad A_0|\mu\rangle=0.\\
 B_0|\sigma\rangle &=0, \qquad B_0|\mu\rangle=\frac{1+i}{\sqrt{2}}|\sigma\rangle.
\end{aligned}
\end{equation}

The above twist fields $\sigma$ and $\mu$ can be  identified with the `spin operators' in the  BMS Ising model. 
Now let us  calculate  the quantum numbers of these twist operators. Since the symmetry algebra is the BMS-Kac-Moody algebra, we need to know  their conformal  dimensions, boost charges  and the $u(1)$ charges. They can be  determined by  considering the expectation value of the current $T$, $M$ and $\cE$ in the R-sector, which can be calculated in two ways: either using the definition of the ground states in the Ramond sector or using the OPE. Comparing the results from these two methods  gives the quantum numbers. We first compute the expectation values of the currents directly. Using the commutation relation of the modes, we find
\begin{equation}\label{R00}
 \langle \psi_0(x_1)\psi_0(x_2)\rangle_\sigma=0, \qquad \langle \psi_0(x_1)\psi_0(x_2)\rangle_\mu=0,
\end{equation}
\begin{equation}\label{R01}
 \langle \psi_0(x_1,y_1)\psi_1(x_2,y_2)\rangle_\sigma=
 \frac{\sqrt{\frac{x_1}{x_2}}}{x_1-x_2}, \quad \langle \psi_0(x_1,y_1)\psi_1(x_2,y_2)
 \rangle_\mu=
 \frac{\sqrt{\frac{x_2}{x_1}}}{x_1-x_2},
\end{equation}
\begin{equation}\label{R10}
    \langle \psi_1(x_1,y_1)\psi_0(x_2,y_2)
 \rangle_\sigma=
 \frac{\sqrt{\frac{x_2}{x_1}}}{x_1-x_2}, \quad \langle \psi_1(x_1,y_1)\psi_0(x_2,y_2)
 \rangle_\mu=
 \frac{\sqrt{\frac{x_1}{x_2}}}{x_1-x_2},
\end{equation}
so in both of the Ramond ground states, we have
\begin{equation}\label{R010}
 \langle \psi_0(x_1,y_1)\psi_1(x_2,y_2)+\psi_1(x_1,y_1)\psi_0(x_2,y_2)
 \rangle_{\sigma/\mu}=\frac{\sqrt{\frac{x_1}{x_2}}+\sqrt{\frac{x_2}{x_1}}}{x_1-x_2}.
\end{equation}
The short distance
behaviour of the above correlators coincides with the corresponding ones in the NS sector, because  short-distance behavior is independent of the global boundary conditions. From their definitions, the currents can be realized as:
\bea
 T(x,y)&=&-\frac{1}{2}\left[\psi_1(z,w)\partial_z\psi_0(x,y)+\psi_0(z,w)\partial_z\psi_1(x,y)+\frac{2}{(z-x)^2}\right]_{z\to x, w\to y},\\
M(x)&=&-\frac{1}{2}\left[\psi_0(z)\partial_z\psi_0(x)\right]_{z\to x},\\
\cE(x,y)&=&-\frac{1}{2}\left[\psi_0(z,w)\psi_1(x,y)-\frac{1}{z-x}\right]_{z\to x, w\to y}.
\eea 
Taking the $z-x=\epsilon\to 0$ limit and using \eqref{R00}-\eqref{R010},  we find the following expectation values:
\begin{equation}\label{expect}
   \langle M(x) \rangle_{\sigma/\mu}=0, \qquad
   \langle T(x,y) \rangle_{\sigma/\mu}=\frac{1}{8x^2}, \qquad
   \langle \cE(x,y) \rangle_\sigma=-\frac{1}{4x}, \qquad
   \langle \cE(x,y) \rangle_\mu=\frac{1}{4x}.
\end{equation}

Next we consider the OPEs between the currents with the spin operators. 
Because $\sigma$ and $\mu$ are both  BMS-Kac-Moody singlet primaries, their OPEs with the  stress tensors and the $u(1)$ current are
\begin{equation}\label{OPEtwi}
\begin{aligned}
 T(x,y)\sigma(0,0)|0\rangle &\sim \frac{\Delta_\s \sigma(0,0)}{x^2}|0\rangle+\frac{2y\xi_\s \sigma(0,0)}{x^3}|0\rangle+..., \\
 M(x,y)\sigma(0,0)|0\rangle &\sim \frac{\xi_\s \sigma(0,0)}{x^2}|0\rangle+...,\\
 \cE(x,y)\sigma(0,0)|0\rangle&\sim \frac{-\varepsilon\sigma(0,0)}{x}|0\rangle+..., \\
 &\mbox{(similar ones for $\mu$)}. 
\end{aligned}
\end{equation}
 From the above OPEs \eqref{OPEtwi},  we have
\begin{equation}
\begin{aligned}
    \langle & M(x) \rangle_{\sigma}=\frac{\xi_\sigma}{x^2}, \qquad
   \langle T(x,y) \rangle_{\sigma}=\frac{\Delta_\sigma}{x^2}, \qquad
   \langle \cE(x,y) \rangle_\sigma=-\frac{\varepsilon_\sigma}{x},\\
   \langle & M(x) \rangle_{\mu}=\frac{\xi_\mu}{x^2}, \qquad
   \langle T(x,y) \rangle_{\mu}=\frac{\Delta_\mu}{x^2}, \qquad
   \langle \cE(x,y) \rangle_\mu=-\frac{\varepsilon_\mu}{x}.
\end{aligned}
\end{equation}
Comparing with \eqref{expect}, we read the quantum number of the twist fields
\begin{equation}\label{twistquan}
 \Delta_\sigma=\Delta_\mu=\frac{1}{8}, \qquad \xi_\sigma=\xi_\mu=0, \qquad
 \varepsilon_\sigma=\frac{1}{4}, \qquad \varepsilon_\mu=-\frac{1}{4}.
\end{equation}
We had obtained these quantum numbers from the representation theory of the zero modes of $\mathfrak{S}$, below \eqref{M0K0}. If we view the twist fields as  $\mathfrak{S}$-primaries, the corresponding $K_0$ charge $\kappa$ is $\kappa=0$, which can also be similarly obtained as $\Delta$, $\xi$ and $\varepsilon$. Note that the values of $\Delta$ and $\xi$ agree with the ones  by taking the non-relativistic limit of the Ising model.

All the states in the Ramond sector can be categorized according to their fermionic numbers. They must lie in the two BMS-Kac-Moody staggered module with singlet ground  states (BMS-Kac-Moody primaries) $|\sigma\rangle$ or $|\mu\rangle$.  Alternatively, from the perspective of the representation of $\mathfrak{S}$, these states belong to one of two  highest-weight modules of $\mathfrak{S}$,  in which $|\sigma\rangle$ and $|\mu\rangle$ are both $\mathfrak{S}$-primaries. Explicitly, we have the following classification:
\begin{itemize}
  \item Module containing $\sigma$: even number of fermions built on $|\sigma\rangle$ $+$ odd number of fermions built on $|\mu\rangle$. They all have $(-1)^F=1$ and the ground state is $|\sigma\rangle$.
  \item Module containing $\mu$: odd number of fermions built on $|\sigma\rangle$ $+$ even number of fermions built on $|\mu\rangle$. They all have $(-1)^F=-1$ and the ground state is $|\mu\rangle$.
\end{itemize}

\subsubsection*{Fusion rules}

Next we turn to determine the fusion rules. The criterion  is to see whether the related three-point function vanish or not.  For the BMS primaries, they  fuse as 
\begin{equation}\label{fusion}
\begin{matrix}
   [\cE][\cE]=[\mathbf{1}], &  [\Psi][\Psi]=[\mathbf{1}]+ [\cE],&  \\
   [\sigma][\sigma]=[\mathbf{1}]+[\cE]& [\mu][\mu]=[\mathbf{1}]+[\cE], & [\sigma][\mu]=[\Psi],\\
   [\sigma][\cE]=[\sigma], & [\mu][\cE]=[\mu], & \\
   [\Psi][\sigma]=[\mu], & [\Psi][\mu]=[\sigma], & [\Psi][\cE]=[\Psi].
   \end{matrix}
\end{equation}
Here we denote the doublet representation of fundamental fields $(\psi_0, \psi_1)$ as $\Psi$. Notice that  the fusion of $[\Psi][\Psi]$ can be understood as follows:  the  chiral and anti-chiral free fermions combine into one  module  of  the doublet of the BMS free fermions,
\begin{equation}
    \psi, \quad \bar{\psi}\qquad \Rightarrow \qquad (\psi_0,  \psi_1), 
\end{equation}
so the corresponding fusion rules can be translated in terms of $\Psi$, 
\begin{equation}
    [\psi][\psi]=[\bar{\psi}][\bar{\psi}]=[\mathbf{1}], \quad [\psi][\bar{\psi}]=[\cE]
    \qquad \Rightarrow \qquad
    [\Psi][\Psi]=[\mathbf{1}]+ [\cE].
\end{equation}
%this point also indicates the important fact that $\mathbf{1}$ and $\epsilon$ should be in one module of the underlying symmetry (CBMS), which  has been shown explicitly in the last  subsection.
%\ZF{}{agree with the revision here.}
The fusion rules \eqref{fusion} form an algebra. It has a closed subalgebra including only $\mathbf{1}$, $\sigma$ and $\cE$, or equivalently  including $\mathbf{1}$, $\mu$ and $\cE$, with the following fusion rules:
\begin{equation}
\begin{matrix}
  [\cE][\cE]=[\mathbf{1}], & [\sigma][\sigma]=[\mathbf{1}]+[\cE]
   & [\sigma][\cE]=[\sigma],\\
   [\mathbf{1}][\cE]=[\cE], & [\mathbf{1}][\sigma]=[\sigma], & [\mathbf{1}][\mathbf{1}]=[\mathbf{1}].
\end{matrix}
\end{equation}
This fusion algebra as well as the original one \eqref{fusion} only concern the BMS primaries and their BMS descendents, so they do not cover all the states in the theory. To include all  states, we need to organize the fusions in terms of BMS-Kac-Moody staggered modules or  highest-weight modules of $\mathfrak{S}$. Therefore the above  fusion rules  can be rewritten more suitably as follows
\begin{equation}\label{1sigma}
    [\mathbf{1}_{\mathfrak{S}}][\mathbf{1}_{\mathfrak{S}}]=[\mathbf{1}_{\mathfrak{S}}], \quad [\mathbf{1}_{\mathfrak{S}}][\sigma_{\mathfrak{S}}]=[\sigma_{\mathfrak{S}}], \quad [\sigma_{\mathfrak{S}}][\sigma_{\mathfrak{S}}]=[\mathbf{1}_{\mathfrak{S}}]
\end{equation}
where the subscript ``$\mathfrak{S}$''  means that they are  $\mathfrak{S}$-modules.
These are  the operator spectrum ($\mathfrak{S}$-primary operators) and the fusion rules for the BMS Ising model.

Next we consider the structure constants in BMS Ising model. Recall that Ising model has one non-trivial structure constant: $C_{\sigma\sigma\epsilon}=\frac{1}{2}$. Interestingly, the BMS Ising model with the underlying algebra be $\mathfrak{S}$,  has no non-trivial structure constant, as can be seen from the above fusion rules. All information of the 4-point correlation function are encoded in the $\mathfrak{S}$-conformal blocks.

\subsection{The partition function}

Now we discuss the partition function. Firstly, let us briefly review the   modular invariance of the partition function in general BMSFTs. The partition function is defined as\footnote{We follow the notation in \cite{Bagchi:2019unf}. Under the identification $\sigma=ia$, $\rho=ib$, our notation  coincides with the one used in \cite{Hao:2021urq}.}:
\begin{equation}\label{partition}
     Z(\sigma,\rho)\equiv \text{Tr}_{\mathcal{H}}e^{2\pi i \sigma(L_0-\frac{c_L}{24})+2\pi i\rho(M_0-\frac{c_M}{24})}
\end{equation}
 where $\sigma$, $\rho$ are  modular parameters. It can be obtained from the partition function of a 2d CFT,
  \begin{equation}\label{CFTpar}
     Z(\tau,\bar{\tau})=\text{Tr}_{\mathcal{H}}e^{2\pi i \tau(\mathcal{L}_0-
     \frac{c}{24})-2\pi i
     \bar{\tau}(\bar{\mathcal{L}}_0-\frac{\bar{c}}{24})}
 \end{equation}
 by taking  either NR or UR limit \cite{Bagchi:2019unf}, with the modular parameters being related by
 \begin{equation}\label{NRUR}
 \begin{aligned}
     \text{NR}&: \quad \tau=\sigma+ \epsilon\rho, \qquad \bar{\tau}=-\sigma+\epsilon\rho,\\
     \text{UR}&: \quad \tau=\sigma+ \epsilon\rho, \qquad \bar{\tau}=\sigma-\epsilon\rho.
 \end{aligned}
 \end{equation}

 The BMS modular invariance had been discussed in the literatures, both intrinsically \cite{Barnich:2012xq}\cite{Jiang:2017ecm}\cite{Hao:2021urq} and by taking the  limit \cite{Bagchi:2012xr}\cite{Bagchi:2013qva}\cite{Bagchi:2019unf}. The modular  $\textbf{S}$-transformation is:
\begin{equation}
   \textbf{S}: \qquad \sigma\rightarrow -\frac{1}{\sigma}, \quad \rho\rightarrow
   \frac{\rho}{\sigma^2}.
  \end{equation}
  and the  modular  $\textbf{T}$-transformation is:
  \begin{equation}\label{morT}
   \textbf{T}: \qquad \sigma\rightarrow \sigma+1, \quad \rho\rightarrow \rho.
  \end{equation}

Now we go back to the partition function of the BMS free fermion. It is easy to see that the partition function of the BMS free fermion is simply the NR limit  of the one of the free Majorana fermion. Equivalently, it is the   UR limit of the partition function  of the free Majorana fermion in the flipped vacuum. Furthermore, by the fermion-boson duality, the partition function of the BMS Ising model is simply the NR/UR limit of the partition function  of the Ising model in the highest weight/flipped vacuum. Thus making use of  \eqref{NRUR}, one can easily obtain the  partition function: 
\begin{equation}\label{par}
  \begin{aligned}
    Z(\sigma,\rho)&\equiv \text{Tr}_{\mathcal{H}}e^{2\pi i \sigma(L_0-\frac{c_L}{24})+2\pi i\rho(M_0-\frac{c_M}{24})}\\
    &=\frac{\theta_2(\sigma)+\theta_3(\sigma)+\theta_4(\sigma)}{2\eta(\sigma)}\\
    &=\chi_0^2(\sigma)+\chi_{\frac{1}{2}}^2(\sigma)+\chi_{\frac{1}{16}}^2(\sigma)=\chi^{(\mathfrak{S})}_0(\sigma)+\chi^{(\mathfrak{S})}_{\frac{1}{8}}(\sigma)
  \end{aligned}
  \end{equation}
  where  $\chi^{(\mathfrak{S})}_\Delta$ denote the character of the $\mathfrak{S}$-module and
  \begin{equation}
  \chi^{(\mathfrak{S})}_0(\sigma)=\frac{\theta_3(\sigma)+\theta_4(\sigma)}{2\eta(\sigma)},\qquad \chi^{(\mathfrak{S})}_{\frac{1}{8}}(\sigma)=\frac{\theta_2(\sigma)}{2\eta(\sigma)}.
  \end{equation}
One can easily check that the partition function \eqref{par} is invariant under the  BMS modular  $\textbf{S}$-transformation\footnote{It is clear that the partition function \eqref{par} is not invariant under the modular  $\textbf{T}$- transformation \eqref{morT}. This suggest that the BMS free fermion is not well-defined on the torus. This is in fact also the case for the BMS free scalar\cite{Hao:2021urq}. Nevertheless, they are consistent theories on the plane.}.

The above partition function can also be viewed as part of a chiral  minimal model based on the $W$-algebra $\mathfrak{S}$. In fact, the full local  minimal model of  $\mathfrak{S}$ is just the free boson compactified on a $S^1$ with unit radius $r=1$. This rational CFT has the following partition function:
\begin{equation}
 Z(\tau,\bar{\tau})_{r=1}=\frac{1}{2}\left(\left|\frac{\theta_2}{\eta}\right|^2+\left|\frac{\theta_3}{\eta}\right|^2+\left|\frac{\theta_4}{\eta}\right|^2\right)=2\chi^{(\mathfrak{S})}_{\frac{1}{8}}\bar{\chi}^{(\mathfrak{S})}_{\frac{1}{8}}+\chi^{(\mathfrak{S})}_0
 \bar{\chi}^{(\mathfrak{S})}_0+\chi^{(\mathfrak{S})}_{\frac{1}{2}}\bar{\chi}^{(\mathfrak{S})}_{\frac{1}{2}}
\end{equation}
where $\chi^{(\mathfrak{S})}_0$ and $\chi^{(\mathfrak{S})}_{\frac{1}{8}}$ are as above, and
\begin{equation}
\chi^{(\mathfrak{S})}_{\frac{1}{2}}=\frac{\theta_3(\sigma)-\theta_4(\sigma)}{2\eta(\sigma)}.
 \end{equation}
 Note that, by the fermion-boson duality, this is also the partition function of the Dirac fermion\cite{Ginsparg:1988ui}\cite{DiFrancesco:1997nk}. The corresponding chiral theory includes three $\mathfrak{S}$-primaries $\mathbf{1}_{\mathfrak{S}}$, $\sigma_{\mathfrak{S}}$ and $\epsilon_{\mathfrak{S}}$ (the Kac spectrum of $\mathfrak{S}$), with dimensions
 \begin{equation}
 \Delta_{\mathbf{1}_{\mathfrak{S}}}=0, \quad \Delta_{\sigma_{\mathfrak{S}}}=\frac{1}{8}, \quad \Delta_{\epsilon_{\mathfrak{S}}}=\frac{1}{2}.
 \end{equation}
 Note that we use the same notation $\mathbf{1}_{\mathfrak{S}}$ and $\sigma_{\mathfrak{S}}$ as in \eqref{1sigma} since they are exactly the same $\mathfrak{S}$-modules.
  This chiral minimal model has the following fusion rules (omit the trivial ones):
\begin{equation}
 [\sigma_{\mathfrak{S}}][\sigma_{\mathfrak{S}}]=[\mathbf{1}_{\mathfrak{S}}],\quad [\epsilon_{\mathfrak{S}}][\epsilon_{\mathfrak{S}}]=[\mathbf{1}_{\mathfrak{S}}], \quad [\epsilon_{\mathfrak{S}}][\sigma_{\mathfrak{S}}]=[\sigma_{\mathfrak{S}}].
\end{equation}
One can see that the fusion algebra of the BMS Ising model, which only involves $\mathbf{1}_{\mathfrak{S}}$ and $\sigma_{\mathfrak{S}}$, is just a fusion subalgebra of this chiral $\mathfrak{S}$-minimal model. This means only the subset $\{\mathbf{1}_{\mathfrak{S}},\sigma_{\mathfrak{S}}\}$ of the Kac spectrum of $\mathfrak{S}$  appears in the BMS Ising model.  Thus the BMS Ising model gives a novel minimal model realization of the underlying algebra.

We would like to comment on the  NR limit of the 2d Ising model. Recall that we have the 
 following relation for the partition function\footnote{In the following, when we say the NR limit, we always means that  there is an alternative choice of the UR limit of the CFT in the flipped vacuum.  In the case at hand, they give rise to the same spectrum as well as the same partition function.}:
\begin{equation}\label{NRIsing}
  \text{Ising model}\xrightarrow[]{\text{NR limit}}\text{BMS Ising}
\end{equation}
or by the fermion-boson duality
\begin{equation}
  \text{Majorana fermion} \xrightarrow[]{\text{NR limit}} \text{BMS free fermion}
\end{equation}
While the partition function of the BMS Ising model is exactly the NR limit of the partition function of the Ising model, the details  are somehow  
puzzling. Firstly, if we naively take the NR limit of the Ising model,  we find that the spectrum includes three BMS primaries:
\begin{equation}
    \mathbf{1} \xrightarrow[]{\text{NR limit}} \mathbf{1}', \quad \sigma
     \xrightarrow[]{\text{NR limit}} \sigma', \quad \epsilon \xrightarrow[]{\text{NR limit}}
     \epsilon' \label{NRspectrum}
\end{equation}
with dimensions and boost charges respectively 
\begin{equation}
    \Delta_{\mathbf{1}'}=0, \quad \xi_{\mathbf{1}'}=0, \quad \Delta_{\sigma'}=\frac{1}{8}, 
    \quad \xi_{\sigma'}=0, \quad \Delta_{\epsilon'}=1, \quad \xi_{\epsilon'}=0.
\end{equation}
One can identify these states with the one in the free field realization as\footnote{We add a  subscripts ``$f$'' to stress that they are obtained in the free field realization.}: $\mathbf{1}'\sim \mathbf{1}_f$, $\sigma'\sim \sigma_f$, $\epsilon'\sim \cE_f$. However, 
 these states, together with their BMS descendants, do not cover all the states we find in the free field realization. More precisely, the operator $K$ can not be read from the NR limit of the Ising model. In fact, under the NR limit, the central charge  become $c_L=c+\bar{c}=1$, $c_M=\epsilon(c-\bar{c})=0$, and the theory reduces simply to a chiral one\cite{Bagchi:2009pe}\cite{Hao:2021urq}. However, the above spectrum \eqref{NRspectrum} is not the one in a chiral minimal model. Secondly, while we can get the fusion rules from the NR limit:
 \begin{equation}
\begin{matrix}
  [\epsilon'][\epsilon']=[\mathbf{1}], & [\sigma'][\sigma']=[\mathbf{1}]+[\epsilon'],
   & [\sigma'][\epsilon']=[\sigma'],\\
   [\mathbf{1}'][\epsilon']=[\epsilon'], & [\mathbf{1}'][\sigma']=[\sigma'], & [\mathbf{1}'][\mathbf{1}']=[\mathbf{1}'].
\end{matrix}
\end{equation}
 the corresponding null condition can not simply be seen from the NR limit.   Finally, in the free field realization, we have shown that the BMS highest-weight module containing $\mathbf{1}'$ ($\mathbf{1}_f$) and the one containing $\epsilon'$ ($\cE_f$) are related to each other, this fact can not be seen from the NR limit point of view either. These three puzzles can all be resolved by the enlarged symmetry and the enlarged module we find in the free field realization. Thus, the free field realization help us to see how the NR limit of the Ising model makes sense and what is the underlying structure: the underlying algebra changes under the NR limit as:
\begin{equation}
  \text{Vir}\times\text{Vir}\rightarrow \mathfrak{S}
\end{equation}
Under the NR limit, such  a change of the underlying algebra may or may not appear in other minimal models.

%Having this picture in mind, one may try to construct possible BMS minimal models  based on other chiral    minimal models. For example, it is interesting to see whether 

It is worthy to emphasize  that the BMS Ising model constructed in this section is not a minimal model with respect to the BMS algebra. In fact, the minimal model construction based on the BMS Kac determinant simply reduce to chiral Virasoro minimal models. We  show this fact in appendix \ref{BMSKac}.

%}{}

\section{Conclusion and outlook}
In this work, we studied a new kind of free BMS field theories called the (inhomogeneous) BMS free fermion, and then used it to give a free field construction of the BMS Ising model. 
The BMS  free fermion exhibit many novel features. Firstly, the underlying symmetry is generated by a new type of BMS-Kac-Moody algebra, which is different from those obtained by contractions. In particular, it includes an anisotropic scaling symmetry so the theory has two different scaling symmetries. It is interesting that such a symmetry also appear in the BMS free scalar model  (see appendix \ref{Aniscalar}). Note that the  Kac-Moody current in our case has a non-vanishing level so  BMS highest-weight modules are enlarged by this current. This is  different from the Kac-Moody current studied in the  \cite{Hao:2021urq}. 

Secondly, the module in the BMS free fermion turns out to be  BMS-Kac-Moody staggered module, which are similar with the BMS staggered module studied in the BMS free scalar model\cite{Hao:2021urq}. In the  staggered vacuum module, an extra operator $K$ appears. Unlike the one in the BMS free scalar, here $K$ is not a BMS quasi-primary. As a result, the Hilbert space can not be organized by the BMS quasi-primaries and their descendents. Relatedly, the OPEs can  not be organized in terms of  BMS quasi-primaries either. Nevertheless, we  showed  that the OPEs  can actually be organized by the Virasoro 
quasi-primaries, which helped us to read an underlying $W$-algebra $\mathfrak{S}$ of the vacuum module. The  BMS-Kac-Moody staggered modules appearing in the BMS free fermion can be viewed as  highest weight modules of $\mathfrak{S}$.

Finally, we studied the Ramond sector to obtain the twist operators, which can be identified with the ``spin operators''  in the BMS Ising model. Making use of this free fermion realization, we obtain the fusion rules of the BMS Ising model. While the partition function of the BMS Ising model is exactly the NR limit of the  partition function of the Ising model, one can not  obtain the BMS Ising model  simply by   taking the  NR limit. We stress that the BMS Ising model is not a BMS minimal model. Instead, the underlying algebra is the enlarged BMS algebra $\mathfrak{S}$, and the BMS Ising model in fact gives a novel minimal model realization of this $W$-algebra.

There are several future directions. Firstly, the anisotropic scaling symmetry deserves to be further explored. It would be nice to do a detailed analysis of this symmetry in the BMS free scalar model. It is also possible to find this symmetry in other BMSFTs, may be other free theories like the ghost system. It will be interesting if one can find such a symmetry in higher dimensional GCFTs or CCFTs, possibly in the free theories. From the holographic point of view, this anisotropic scaling corresponds to the ``super-diltation'' in the conformal BMS algebra in the bulk \cite{Fuentealba:2020zkf}\cite{Haco:2017ekf}.

It would be interesting to see whether there are other BMS models which could be minimal with respect to some enlarged BMS algebras. For example, as $\mathfrak{S}$ is the quantum conformal BMS algebra with $c=1$,   one can try to study  generic quantum conformal BMS algebra and search for other possible minimal models. One can also discuss enlarged  super-symmetric BMS algebras and the corresponding minimal models. An explicit example (of the algebra) is constructed in \cite{Fuentealba:2020zkf}, which is the super-symmetric version of the classical conformal BMS$_3$ algebra. This is a $W$-algebra of type $W(2,2,2,\frac{3}{2}, \frac{3}{2}, 1)$.  It is possible that one can construct  the super-symmetric BMS Ising model based on this super conformal BMS$_3$ algebra. A free field realization may be helpful as well in this case.

It will be interesting to see whether the BMS free fermion and the BMS Ising model can be related to the BMS scalar model. It is not clear how to do the bosonization in the BMS case. In particular,  it is not clear how to develop the Coulomb gas formalism in the BMS case, which may be helpful to study the BMS Ising model as well as other possible minimal models.

\section*{Acknowledgments}
    We are grateful to  Chi-ming Chang, Peng-xiang Hao, Reiko Liu, Wei Song, Yu-fan Zheng for valuable discussions, and to Daniel Grumiller and Iva Lovrekovic for useful correspondence. We would like to thank the participants in the "Third National Workshop on Quantum Fields and String" (Beijing 2022-08) for stimulating discussions. The work is supported in part by NSFC Grant  No. 11735001.

\appendix

\section{Anisotropic scaling symmetry in the BMS free scalar }\label{Aniscalar}
In this section, we show that a similar anisotropic scaling symmetry also appear in the BMS free scalar model. The BMS free scalar model has the action:
\begin{equation}\label{BMSscalar}
    S=-\frac{1}{2}\int dxdy (\partial_0\phi)^2.
\end{equation}
In \cite{Hao:2021urq}, it was shown that the symmetry of  this theory is a BMS-Kac-Moody algebra, including the BMS symmetry as well as a $u(1)$ affine current symmetry coming from the $y$-independent translation of the fundamental scalar: 
\begin{equation}
    \phi\to \phi+\Lambda(x).
\end{equation}
In fact, the action \eqref{BMSscalar} has another scaling symmetry, which is anisotropic
\begin{equation}
    \mathfrak{D}': \quad x\to x, \quad y\to\lambda y 
\end{equation}
with the fundamental field transforming as
\begin{equation}
    \phi\to \lambda^{\frac{1}{2}}\phi.
\end{equation}
The corresponding Noether current is
\begin{equation}
\begin{aligned}
      J^{0}_{\mathfrak{D}'}&=-\frac{1}{2}\partial_0\phi\partial_0\phi y+
      \frac{1}{2}\phi\partial_0\phi= M(x)y-\mathfrak{J}(x,y),\\
      J^1_{\mathfrak{D}'}&=0,
\end{aligned}
\end{equation}
where $M(x)=-\frac{1}{2}\partial_0\phi\partial_0\phi$ is the stress tensor of the model\cite{Hao:2021urq}, and 
we has introduced the current
\begin{equation}
    \mathfrak{J}(x,y)\equiv -\frac{1}{2}\phi\partial_0\phi.
\end{equation}
From the equations of motion $\partial_0^2\phi=0$, it is clear that 
$\partial_0^2\mathfrak{J}=0$, and we have
\begin{equation}
    \mathfrak{J}(x,y)=\mathfrak{J}^{(0)}(x)+\mathfrak{J}^{(1)}(x)y.
\end{equation}
The Noether current is conserved
\begin{equation}
    \partial_\mu J^{\mu}_{\mathfrak{D}'}=0,
\end{equation}
which indicates
\begin{equation}\label{scalarJM}
    \partial_0\mathfrak{J}=M.
\end{equation}
i.e. $\mathfrak{J}^{(1)}(x)=M$, and 
\begin{equation}
    J^{0}_{\mathfrak{D}'}=-\mathfrak{J}^{(0)}(x).
\end{equation}
We denote $J^{0}_{\mathfrak{D}'}$ as $\tilde{J}$. One can see that the above analysis is quite similar with the one in the BMS free fermion. Similarly,  we can also find another scaling symmetry (denote the scaling in the BMS algebra as $\mathfrak{D}$):
\begin{equation}
\begin{aligned}
    \mathfrak{D}'': \quad x&\to \lambda x, \quad y\to y\\
      \phi&\to\lambda^{-\frac{1}{2}}\phi
\end{aligned}
\end{equation}
which is not an independent one since $\mathfrak{D}_\lambda''=\mathfrak{D}_{\lambda^{-1}}'\mathfrak{D}_\lambda$.

Similar to the BMS free fermion, the scaling symmetry $\mathfrak{D}'$ can be enhanced to a local symmetry (with $\mathfrak{D}'= \mathfrak{C}_\lambda$)
\begin{equation}
\begin{aligned}
      \mathfrak{C}_f: \quad x&\to x, \quad y\to f(x)y,\\
      \phi &\to f^{\frac{1}{2}}\phi,
\end{aligned}
\end{equation}
 leading to an affine $u(1)$ symmetry with the charges $\tilde{J}_n$, which are the modes of $\tilde{J}$. The generators of $\mathfrak{C}_f$ (denoted as $\tilde{j}_n$) is the same as $j_n$ for $\mathcal{C}_f$ in \eqref{Jvector}, so from $\tilde{j}_n, l_n, m_n$ we can obtain the same BMS-Kac-Moody algebra \eqref{ClaBMSK} without central extensions.  One can compute the OPE of the currents to find the corresponding quantum algebra with central extensions. While it is interesting,  further  discussions on this symmetry in the BMS free scalar model will be beyond the scope of present work.

\section{Bottom up construction of enlarged BMS algebras}\label{boo}

%\ZF{}{The appendix B.1 is completed  so please check and revise this part firstly. }

In this section,  we will study  enlarged BMS algebras from a bottom up construction. Firstly, we study general BMS-Kac-Moody algebra with a $u(1)$ Kac-Moody current. In particular, we manage to  determine all the possible central extensions of the BMS-Kac-Moody algebra \eqref{ClaBMSK}. We  find that  the central charge 
$c_M$ must be vanishing while $c_L$ and the Kac-Moody level $k$  can be arbitrary.   Then, we show that when this BMS-Kac-Moody algebra is further enlarged to the quantum conformal BMS algebra (a W-algebra of type $W(2,2,2,1)$) \cite{Fuentealba:2020zkf},  the relation of the  central charge $c$ and the Kac-Moody level $k$ will be determined by the Jacobi identities. In particular, the algebra $\mathfrak{S}$ is  the quantum conformal BMS algebra with $k=\frac{1}{4},\, c=1$.

\subsection{BMS-Kac-Moody algebra}\label{classifaction}
We will study  general  BMS-Kac-Moody algebras with a $u(1)$ Kac-Moody current.  We start with the BMS$_3$ algebra:
\begin{equation}
\begin{aligned}
  \left[L_n, L_m\right]&=(n-m)L_{n+m}+\frac{c_L}{12}n(n^2-1)\delta_{n+m,0},\\
  [L_n, M_m]&=(n-m)M_{n+m}+\frac{c_M}{12}n(n^2-1)\delta_{n+m,0},\\
  [M_n, M_m]&=0,
\end{aligned}
\end{equation}
then we include a current of dimension 1 into the algebra, which we denote by\footnote{Here we use the same notation $J$ as the $u(1)$ current in the BMS free fermion theory.} $J$. By definition, the current has the following commutation relation with $L$
\begin{equation}
    [L_n, J_m]=-mJ_{n+m}
\end{equation}
and
\begin{equation}
    [J_n, J_m]=kn\delta_{n+m,0}
\end{equation}
where $k$ is the Kac-Moody level.
To determine the last commutator $[M_n, J_m]$, we need the general formula for the  commutator of the modes of two quasi-primaries \cite{Blumenhagen:1990jv}:
\begin{equation}\label{twoquasi}
[\phi_{(i)n},\phi_{(j)m}]=\sum_kC_{ij}^kp_{ijk}\phi_{(k)n+m}+d_{ij}\delta_{n+m,0}
\left(\begin{matrix}
  n+h_i-1 \\
   2h_i-1 \end{matrix}\right)
\end{equation}
where
\begin{equation}
    p_{ijk}(n,m)=\sum_{\begin{matrix}
   r,s\in \mathbb{N} \\
   r+s=h_i+h_j-h_k-1\end{matrix}}C^{ijk}_{r,s}\left(\begin{matrix}
  -n+h_i-1 \\
   r \end{matrix}\right)
  \left(\begin{matrix}
  -m+h_j-1 \\
  s \end{matrix}\right)
\end{equation}
and
\begin{equation}
    C_{r,s}^{ijk}=(-1)^r\frac{(2h_k-1)!}{(h_i+h_j+h_k-2)!}\prod_{t=0}^{s-1}(2h_i-r-t-2)\prod_{u=0}^{r-1}
    (2h_j-s-u-2).
\end{equation}
According to this formula, $p_{212}(n,m)=1$ and $p_{211}(n,m)=-m$, so
the commutation relation between $M$  and $J$ could be:
\begin{equation}\label{MJgeneral}
    [M_n, J_m]=C_{MJ}^LL_{n+m}+C_{MJ}^MM_{n+m}+
    C_{MJ}^{\hat{N}(JJ)}\hat{N}(JJ)_{n+m}-C_{MJ}^J mJ_{n+m}.
\end{equation}
 Then considering the Jacobi identity:
\begin{equation}\label{jacobi1}
    [[M_n,M_m],J_p]+[[M_m,J_p],M_n]+[[J_p,M_n],M_m]=0,
\end{equation}
and focusing on the $J_{n+m}$ terms, we find
\begin{equation}
    C_{MJ}^J=0,
\end{equation}
 and we have $\langle M\hat{N}(JJ)\rangle=0$. We can diagonalize the quasi-primaries as
 \begin{equation}\label{diago}
\begin{aligned}
    &\Omega(JJ)\equiv \hat{N}(JJ)-\frac{2k}{c_L}L\\
    &\tilde{M}\equiv M+\frac{c_M}{1-c_L}L-\frac{c_M}{2k(1-c_L)}\hat{N}(JJ)
\end{aligned}
\end{equation}
so that the two point functions are
\begin{equation}
     \langle L \Omega(JJ)\rangle=\langle L \tilde{M}\rangle=
     \langle \Omega(JJ)  \tilde{M}\rangle=0
\end{equation}
Then only the following term remains in  \eqref{MJgeneral},
\begin{equation}
    [M_n, J_m]=C_{MJ}^{\tilde{M}}\tilde{M}_{n+m}.
\end{equation}
Focusing on the constant terms in \eqref{jacobi1}, we read
\begin{equation}\label{2cases}
    C_{MJ}^{\tilde{M}}=0, \quad \text{or} \quad c_M=0.
\end{equation}
All other Jacobi identities give no extra constraint,  so we conclude that there are the following two cases.
\begin{itemize}
    \item Type 1: when  $C_{MJ}^{\tilde{M}}=0$, there is
          \begin{equation}
             [M_n, J_m]=0, 
          \end{equation}  
         and $c_L$, $c_M$ and $k$ can be arbitrary. When $c_L=1$, $c_M=0$, $k=0$, this is the  
         BMS-Kac-Moody algebra in the BMS free scalar model \cite{Hao:2021urq}.
  \item Type 2: when $c_M=0$ and $C_{MJ}^{\tilde{M}}\neq 0$, 
         \begin{equation}\label{type2}
              [M_n, J_m]=-M_{n+m}
         \end{equation}
         where we normalize\footnote{We can do this because we keep the level $k$ undetermined. This normalization is the one used for $\mathfrak{S}$. } $C_{MJ}^{\tilde{M}}=-1$, and $c_L$ and $k$ can be arbitrary. This case gives the possible central extension in the algebra \eqref{ClaBMSK}. When $c_L=1$,  $k=\frac{1}{4}$, this is the  BMS-Kac-Moody algebra in the BMS free fermion.
\end{itemize}
Besides these two cases,  there are other possibilities corresponding to singular cases in \eqref{diago}: $c_L=0$ or  $c_L=1$ or $k=0$. We refer to these singular cases as Type 3. For example, we have:
\begin{itemize}
    \item Type 3: when $k=0$, $J$ and $N(JJ)$ are null states, so we have the following possibility:
    \begin{equation}
         [M_n, J_m]=\alpha N(JJ)_{n+m}
    \end{equation}
    where $\alpha$ is an arbitrary number. $c_L$ and $c_M$ can be arbitrary.
\end{itemize}
There are other possibilities for $[M_n, J_m]$ in these singular cases (Type 3). However, these cases are not relevant with the quantum conformal BMS algebra discussed in the following. Thus we do not list the rest of them.

%\ZF{}{Appendix B.2 is completed, please check and revise.}

%\ZF{}{In the following, I have corrected some normal ordering symbol (to be $\hat{N}$) as well as other typos.}

\subsection{Quantum conformal BMS algebra}

Now we give an explicit construction of the quantum conformal BMS$_3$ algebra. It is  a $W$-algebra of type $W(2,2,2,1)$\cite{Fuentealba:2020zkf}. We must insert another current $K$ with $\Delta=2$ into the algebra, besides the BMS-Kac-Moody currents discussed above.  The quantum conformal BMS$_3$ algebra can be realized by  the quantum Drinfeld-Sokolov (QDS) reduction as\footnote{For a review of the QDS, see \cite{Bouwknegt:1992wg}. However, \cite{Bouwknegt:1992wg} mainly discuss principal embeddings of $sl_2$. For a complete list of embedding types and the  types of the corresponding (classical) $W$-algebras, see \cite{Frappat:1992bs}. For general QDS, the cohomology was resolved in \cite{deBoer:1993iz}\cite{2003Quantum}. For the case at hand, see \cite{2022CMaPh.390...33F} for  more discussions.}
\begin{equation}\label{emb}
  sl_2 \xrightarrow[\text{embedding}]{\text{a non-principal}} B_2\Rightarrow W(2,2,2,1)
 \end{equation}
Note that the embedding of the $sl_2$ is non-principal. It is well known that  QDS always produces  $W$-algebras with generic $c$.  In the following, we will show that the algebra $\mathfrak{S}$ is the  quantum conformal BMS$_3$ algebra with $k=\frac{1}{4},\,c=1$.

The BMS-Kac-Moody subalgebra of the quantum conformal BMS$_3$ algebra should be of type 2 \eqref{type2} derived in the last subsection. One can know this fact from the classical conformal BMS algebra in \cite{Fuentealba:2020zkf}. This means that there must be $c_M=0$. 
We need to determine the commutators involving $K$. From the  embedding \eqref{emb}, it is easy to see that  the  quantum conformal BMS$_3$ algebra will be invariant under the transformation $J\to -J$, $K\to M$ and $M\to K$. Thus  we have
\begin{equation}
\begin{aligned}
    \left[M_n, M_m\right]=0 \quad &\Rightarrow \quad[K_n, K_m]=0, \\
    [M_n, J_m]=-M_{n+m} \quad &\Rightarrow \quad [K_n, J_m]=K_{n+m},\\
    [L_n, M_m]=(n-m)M_{n+m} \quad &\Rightarrow \quad [L_n, K_m]=(n-m)K_{n+m}.
\end{aligned}
\end{equation}
The last and the most interesting commutator is $[M_n,K_m]$. For this commutator, we first need to know the quasi-primary fields up to $\Delta=3$. They are
\begin{itemize}
    \item For $\Delta=1$: J.
    \item For $\Delta=2$: $L$, $M$, $K$, $\mathcal{N}(JJ)$
    \item For $\Delta=3$: $\mathcal{N}(J L)$, $\mathcal{N}(J M)$, $\mathcal{N}(J K)$, $\mathcal{N}(JJJ)$
\end{itemize}
where
\begin{equation}
\begin{aligned}
  \mathcal{N}(JJ)=\hat{N}(JJ),  \quad \mathcal{N}(J L)=\hat{N}(J L), \quad \mathcal{N}(JJJ)=\hat{N}(JJJ)\\
  \mathcal{N}(J M)=4\hat{N}(J M)-\partial M, \quad
  \mathcal{N}(J K)=4\hat{N}(J K)+\partial K.
\end{aligned}
\end{equation}
Using the known commutators, we can easily find
\begin{equation}
    C_{MK}^M= C_{MK}^K= C_{MK}^{\mathcal{N}(J M)}=C_{MK}^{\mathcal{N}(J K)}=0.
\end{equation}
Next, to simplify the manipulation, we diagonalize the remaining  quasi-primaries as
\begin{equation}
\begin{aligned}
    \Lambda(JJ)&=L-\frac{c}{2k}\hat{N}(JJ),\\
    \Lambda(JJJ)&=\hat{N}(JL)-\frac{c+2}{6k}\hat{N}(JJJ),
\end{aligned}
\end{equation}
such that
\begin{equation}
     \langle L| \Lambda(JJ)\rangle=0, \quad  \langle \hat{N}(J L)|\Lambda(JJJ)\rangle=0
\end{equation}
where $c\equiv c_L$ (we will always use this notation for $c_L$ hereafter). 
Then we normalize the $\langle M|K\rangle$ as
\begin{equation}
    \langle M|K\rangle=\frac{c}{2}.
\end{equation}
After these preparations,  we read 
\begin{equation}
\begin{aligned}
    \left[M_n,K_m\right]&=C_{MK}^Lp_{222}(n,m)L_{n+m}+C_{MK}^{J}p_{221}(n,m)J_{n+m}\\
    &+C_{MK}^{\hat{N}(J L)}p_{223}(n,m)\hat{N}(J L)_{n+m}+C_{MK}^{\Lambda(JJJ)}p_{223}(n,m)\Lambda(JJJ)_{n+m}\\
    &+C_{MK}^{\Lambda(JJ)}p_{222}(n,m)\Lambda(JJ)_{n+m}+\frac{c}{2}\frac{n(n^2-1)}{6}\delta_{n+m,0}
\end{aligned}
\end{equation}
where (according to \eqref{twoquasi})
\begin{equation}
    p_{222}(n,m)=\frac{n-m}{2}, \quad p_{221}(n,m)=\frac{m^2+n^2-nm-1}{6}, \quad p_{223}(n,m)=1.
\end{equation}
Using the known commutators, we can calculate the coefficients, which are
\begin{equation}
    C_{MK}^L=2, \quad C_{MK}^{J}=\frac{c}{2k}, \quad C_{MK}^{\Lambda(JJ)}=\frac{4k-c}{2k(c-1)},
\end{equation}
\begin{equation}
     C_{MK}^{\hat{N}(J L)}=\frac{2c}{k(c+2)}, \quad C_{MK}^{\Lambda(JJJ)}=\frac{c(12k-c-2)}{2k^2(c+2)(c-1)},
\end{equation}
so we obtain the commutator 
\begin{equation}
\begin{aligned}
    \left[M_n,K_m\right]&=(n-m)L_{n+m}+\frac{c(m^2+n^2-nm-1)}{12k}J_{n+m}\\
    &+\frac{2c}{k(c+2)}\hat{N}(J L)_{n+m}+\frac{c(12k-c-2)}{2k^2(c+2)(c-1)}\Lambda(JJJ)_{n+m}\\
   &+\frac{(4k-c)(n-m)}{4k(c-1)}\Lambda(JJ)_{n+m} +\frac{c}{2}\frac{n(n^2-1)}{6}\delta_{n+m,0}.
\end{aligned}
\end{equation}
Now we need to consider the Jacobi identities. Using the following identity
\begin{equation}
  [[M_n,K_m],M_p]+[[K_m,M_p],M_n]+[[M_p,M_n],K_m]=0,
\end{equation}
we find the condition
\begin{equation}\label{ckrelation}
 c=-12k+28-\frac{30}{k+1}.
\end{equation}
One can also consider other Jacobi identities, such as:
\begin{equation}\label{MKJ}
      [[M_n,K_m], J_p]+[[K_m, J_p],M_n]+[[J_p,M_n],K_m]=0.
\end{equation}
However, no additional constraints on $c,k$ could be found from them\footnote{In the previous versions of this work, we made a computational error and claimed an (additional) incorrect relation $12k=4-c$ from the Jacobi identity \eqref{MKJ}. We then combined this relation with \eqref{ckrelation} to conclude that the conformal BMS algebra only exists for  $k=\frac{1}{4},\, c=1$, which is not true. Nevertheless, this computational error did not affect our main results  since we  focus on the algebra with $k=\frac{1}{4},\, c=1$.  We thank Daniel Grumiller and Iva Lovrekovic for bringing this error to our attention through their independent construction of the algebra \cite{Grumiller:2026kgx}.  }.  

Note that in the classical limit  $k,c\to\infty$, the equation \eqref{ckrelation} coincides with the equation ``$c=\tilde{c}$'' in the   classical conformal BMS algebra obtained in \cite{Fuentealba:2020zkf}\footnote{Their notation are related to ours by  $c_{\text{their}}=\frac{1}{12}c_{\text{ours}}$, $\tilde{c}=-k$.}. Besides, now the commutator becomes
\begin{equation}\label{Intermsofc}
\begin{aligned}
    \left[M_n,K_m\right]&=(n-m)L_{n+m}-\frac{(6k^2-8k+1)(m^2+n^2-nm-1)}{6k(k+1)}J_{n+m}\\
    &+\frac{2(6k^2-8k+1)}{3k^2(2k-3)}\hat{N}(J L)_{n+m}-\frac{6k^2-8k+1}{3k^2(k-1)(2k-3)}\Lambda(JJJ)_{n+m}\\
   &-\frac{(2k-1)(n-m)}{6k(k-1)}\Lambda(JJ)_{n+m} -\frac{(6k^2-8k+1)n(n^2-1)}{6(k+1)}\delta_{n+m,0}.
\end{aligned}
\end{equation}
One may verify that, under the replacements
\begin{equation}
  [ , ]\to \frac{i}{\hbar}\{ , \}, \quad c\to \frac{c}{\hbar}, \quad L_n\to 
  \frac{L_n}{\hbar}, \quad M_n\to \frac{M_n}{\hbar}, \quad  K_n\to \frac{K_n}{\hbar}, \quad J_n\to \frac{J_n}{\hbar}
\end{equation}
and in the classical limit $\hbar\to 0$, \eqref{Intermsofc} reduces to the classical conformal BMS algebra derived in \cite{Fuentealba:2020zkf}. Thus, it is reasonable to name this algebra as the quantum conformal BMS algebra. One can find interesting unitary field theories with this symmetry algebra in \cite{Grumiller:2026kgx}.

For the special case $c=1$ and $k=\frac{1}{4}$ (which satisfy \eqref{ckrelation}), the above commutator simplifies much and reduces to 
\begin{equation}\label{botMK}
\begin{aligned}
    \left[M_n,K_m\right]&=2(n-m)\hat{N}(JJ)+\frac{m^2+n^2-nm-1}{3}J_{n+m}\\
    &+\frac{16}{3}\hat{N}(JJJ)_{n+m}+\frac{n(n^2-1)}{12}\delta_{n+m,0}.
\end{aligned}
\end{equation}
One can see that $L$ as well as $\hat{N}(J L)$ decouples from the algebra. In fact, because $c=1$ is identical with the central charge of the one from the $u(1)$ Sugawara construction,
\begin{equation}
    T=\frac{1}{2k}\hat{N}(JJ)=2\hat{N}(JJ)
\end{equation}
so both $\Lambda(JJ)$ and $\Lambda(JJJ)$ are  null states. Modding out these null states, we can effectively identify $L$ with $2\hat{N}(JJ)$ and $\hat{N}(J L)$ with  $2\hat{N}(JJJ)$ in the $W$-algebra. Therefore \eqref{botMK} exactly coincides with \eqref{MKcomplete}, and we  obtain the $W$-algebra $\mathfrak{S}$.

\section{BMS Kac determinant and null states }\label{BMSKac}
A natural way to construct BMS minimal models is to calculate its Kac determinant and find the null states and the fusion rules. In this appendix we  show that this construction must lead to $c_M=0$ and furthermore the decoupling of all $M_n$'s from the BMS module, and the possible BMS minimal models have to be chiral \footnote{The BMS Gram matrix and the Kac determinant had been discussed in literature, e.g. \cite{Bagchi:2009pe}\cite{Bagchi:2019unf}, however, a closed formula for the BMS Kac determinant is absent in literature. We learned the following closed formula from  Peng-xiang Hao.}.

The BMS Kac determinant at level $N$ is given by
\begin{equation}
  \text{det}M_{N}(c_M, \xi)=(-1)^N\left[\prod_{ab\leq N, a,b\in \mathbb{N}^+}\chi(a,b)^{\theta(a,b)}\right]^2
\end{equation}
where
\begin{equation}
  \chi(a,b)=\left(2a\xi+\frac{c_M}{12}a(a^2-1)\right)^bb! ,
\end{equation}
\begin{equation}
  \theta(a,b)=\sum_{i=0}^{N-ab}P(i)f(N-ab-1, a) .
\end{equation}
Here $P(N)$ is the number of partitions of $N$, and $f(N,a)$ is the number of partitions of $N$ with the integer $a$ not appearing in the
partition, which can be determined by the following relation
\begin{equation}
  \sum_{N=0}^{\infty}f(N,a)x^N=\prod_{k\neq a}^{\infty}\frac{1}{1-x^k}.
\end{equation}
Note that the above BMS Kac determinant  depends only on $c_M$ and $\xi$, while there are the elements which have $c_L$ and(or) $\Delta$ dependence in the Gram matrix. These elements do not contribute to the  Kac determinant because they always multiply with zeros. These zeros appear due to the fact that all $M_n$'s commutes with each other.

The null states are determined by the vanishing curve
\begin{equation}
  \chi(a,b)=0. \label{vanish1}
\end{equation}
When these null states are modded out, one can compute the Kac determinant of the Gram matrix of the remaining states. It turns out that this Kac determinant depends on $c_L$ and $\Delta$, and the corresponding vanishing condition  for generic $\xi$ and $c_M$ satisfying \eqref{vanish1} is
\begin{equation}\label{vanish2}
    \Delta+\frac{c_L(a^2-1)}{24}=A(a,b),
\end{equation}
where $A(a,b)$ is some function of $a$ and $b$ and its concrete form is not important to us.
To find potential minimal models, we need first find the intersection points of vanishing curves \eqref{vanish1}, which is
\begin{equation}
  c_M=0, \qquad \xi=0.
\end{equation}
In fact, under this condition, there will be more emergent null states and the theory reduce to a chiral CFT \cite{Bagchi:2009pe}\cite{Hao:2021urq}. So after modding out the null states, the  vanishing condition for the Kac determinant of the remaining states is no longer be the generic one \eqref{vanish2}. Instead, we find the chiral part of standard Virasoro minimal models.
As is well-known, while these chiral minimal models has a closed fusion algebra coming from the null states condition,  they can not be a full local theory. A crucial point is the loss of  modular invariance of their partition functions. % in order to be modular invariant, we  must include its anti-chiral counterpart.}{}

 A natural question is whether there exist non-trivial minimal models  when the underlying BMS symmetry is enlarged. Let us first  review the enlarged BMS algebras or Galilean conformal algebras in the literatures briefly. One way to  enlarge the GCA$_2$  is considering the Galilean (non-relativistic) contractions of various $W$-algebras. The resulting algebra is known as the Galilean $W$-algebras, e.g. \cite{Rasmussen:2017eus}. In contrast, another way to enlarge BMS$_3$ (GCA$_2$) is to consider the ultra-relativistic contraction of various $W$-algebras \cite{Afshar:2013vka}\cite{Campoleoni:2016vsh}. As pointed out in \cite{Campoleoni:2016vsh}, there is an interesting feature regarding these two contractions on non-linear $W$-algebras: while the difference between these two contractions can only be seen
 from the representations for linear  $W$-algebras\footnote{For example, for two Virasoro algebras, the NR and UR contractions lead to the same algebra,  GCA$_2\simeq$ BMS$_3$. One can see the differences only from the representation theory: starting from the Virasoro highest weight representation, after the NR and UR contractions, one arrive at the highest-weight representation and the induced representation of the GCA$_2$ respectively.}, these two contractions  lead to different quantum $W$-algebras when acting on non-linear 
 $W$-algebras, even though the resulting classical $W$-algebras are isomorphic. %will always be isomorphic for the NR and UR contractions.}{}
One may try to construct the minimal models with respect to these NR- or UR-constracted $W$-algebras. However, the situation will be similar with the one for the BMS$_3$ (GCA$_2$): one can not obtain any non-trivial minimal models, other than some  chiral minimal models with respect to the original $W$-algebra, which can not be a full local theory. In particular, the generators $M_n$s will always decouple from the module. Thus, to obtain non-trivial minimal models, one essentially needs to enlarge the module to include a special operator $K$ with dimension $\Delta=2$, which prevents the decoupling of $M_n$. This is the case in the BMS Ising model we constructed.

\begin{refcontext}[sorting=none]
\printbibliography

@article{Maldacena:1997re,
    author = "Maldacena, Juan Martin",
    title = "{The Large N limit of superconformal field theories and supergravity}",
    eprint = "hep-th/9711200",
    archivePrefix = "arXiv",
    reportNumber = "HUTP-97-A097, HUTP-98-A097",
    doi = "10.1023/A:1026654312961",
    journal = "Adv. Theor. Math. Phys.",
    volume = "2",
    pages = "231--252",
    year = "1998"
}

@article{Brown:1986nw,
    author = "Brown, J. David and Henneaux, M.",
    title = "{Central Charges in the Canonical Realization of Asymptotic Symmetries: An Example from Three-Dimensional Gravity}",
    doi = "10.1007/BF01211590",
    journal = "Commun. Math. Phys.",
    volume = "104",
    pages = "207--226",
    year = "1986"
}

@article{Strominger:1997eq,
    author = "Strominger, Andrew",
    title = "{Black hole entropy from near horizon microstates}",
    eprint = "hep-th/9712251",
    archivePrefix = "arXiv",
    reportNumber = "HUTP-97-A106",
    doi = "10.1088/1126-6708/1998/02/009",
    journal = "JHEP",
    volume = "02",
    pages = "009",
    year = "1998"
}

@article{Guica:2008mu,
    author = "Guica, Monica and Hartman, Thomas and Song, Wei and Strominger, Andrew",
    title = "{The Kerr/CFT Correspondence}",
    eprint = "0809.4266",
    archivePrefix = "arXiv",
    primaryClass = "hep-th",
    doi = "10.1103/PhysRevD.80.124008",
    journal = "Phys. Rev. D",
    volume = "80",
    pages = "124008",
    year = "2009"
}

@article{Compere:2013bya,
    author = "Comp\`ere, Geoffrey and Song, Wei and Strominger, Andrew",
    title = "{New Boundary Conditions for AdS3}",
    eprint = "1303.2662",
    archivePrefix = "arXiv",
    primaryClass = "hep-th",
    doi = "10.1007/JHEP05(2013)152",
    journal = "JHEP",
    volume = "05",
    pages = "152",
    year = "2013"
}

@article{Compere:2013aya,
    author = "Comp\`ere, Geoffrey and Song, Wei and Strominger, Andrew",
    title = "{Chiral Liouville Gravity}",
    eprint = "1303.2660",
    archivePrefix = "arXiv",
    primaryClass = "hep-th",
    doi = "10.1007/JHEP05(2013)154",
    journal = "JHEP",
    volume = "05",
    pages = "154",
    year = "2013"
}

@article{Compere:2009zj,
    author = "Compere, Geoffrey and Detournay, Stephane",
    title = "{Boundary conditions for spacelike and timelike warped $AdS_{3}$ spaces in topologically massive gravity}",
    eprint = "0906.1243",
    archivePrefix = "arXiv",
    primaryClass = "hep-th",
    reportNumber = "NSF-KITP-09-96, ESI-2149",
    doi = "10.1088/1126-6708/2009/08/092",
    journal = "JHEP",
    volume = "08",
    pages = "092",
    year = "2009"
}

@article{Bagchi:2010zz,
    author = "Bagchi, Arjun",
    title = "{Correspondence between Asymptotically Flat Spacetimes and Nonrelativistic Conformal Field Theories}",
    eprint = "1006.3354",
    archivePrefix = "arXiv",
    primaryClass = "hep-th",
    doi = "10.1103/PhysRevLett.105.171601",
    journal = "Phys. Rev. Lett.",
    volume = "105",
    pages = "171601",
    year = "2010"
}

@article{Bagchi:2012cy,
	archiveprefix = {arXiv},
	author = {Bagchi, Arjun and Fareghbal, Reza},
	doi = {10.1007/JHEP10(2012)092},
	eprint = {1203.5795},
	journal = {JHEP},
	pages = {092},
	primaryclass = {hep-th},
	reportnumber = {EMPG-12-04, NI-12014},
	title = {{BMS/GCA Redux: Towards Flatspace Holography from Non-Relativistic Symmetries}},
	volume = {10},
	year = {2012}}

@article{Barnich:2012xq,
	archiveprefix = {arXiv},
	author = {Barnich, Glenn},
	doi = {10.1007/JHEP10(2012)095},
	eprint = {1208.4371},
	journal = {JHEP},
	pages = {095},
	primaryclass = {hep-th},
	reportnumber = {ULB-TH-12-14},
	title = {{Entropy of three-dimensional asymptotically flat cosmological solutions}},
	volume = {10},
	year = {2012}}

@article{Bagchi:2012xr,
	archiveprefix = {arXiv},
	author = {Bagchi, Arjun and Detournay, St\'ephane and Fareghbal, Reza and Sim\'on, Joan},
	doi = {10.1103/PhysRevLett.110.141302},
	eprint = {1208.4372},
	journal = {Phys. Rev. Lett.},
	number = {14},
	pages = {141302},
	primaryclass = {hep-th},
	reportnumber = {EMPG-12-18},
	title = {{Holography of 3D Flat Cosmological Horizons}},
	volume = {110},
	year = {2013}}

@article{Bagchi:2014iea,
	archiveprefix = {arXiv},
	author = {Bagchi, Arjun and Basu, Rudranil and Grumiller, Daniel and Riegler, Max},
	doi = {10.1103/PhysRevLett.114.111602},
	eprint = {1410.4089},
	journal = {Phys. Rev. Lett.},
	number = {11},
	pages = {111602},
	primaryclass = {hep-th},
	reportnumber = {TUW-14-14},
	title = {{Entanglement entropy in Galilean conformal field theories and flat holography}},
	volume = {114},
	year = {2015}}

@article{Jiang:2017ecm,
	archiveprefix = {arXiv},
	author = {Jiang, Hongliang and Song, Wei and Wen, Qiang},
	doi = {10.1007/JHEP07(2017)142},
	eprint = {1706.07552},
	journal = {JHEP},
	pages = {142},
	primaryclass = {hep-th},
	title = {{Entanglement Entropy in Flat Holography}},
	volume = {07},
	year = {2017}}

@article{Hijano:2018nhq,
    author = "Hijano, Eliot",
    title = "{Semi-classical BMS$_{3}$ blocks and flat holography}",
    eprint = "1805.00949",
    archivePrefix = "arXiv",
    primaryClass = "hep-th",
    doi = "10.1007/JHEP10(2018)044",
    journal = "JHEP",
    volume = "10",
    pages = "044",
    year = "2018"
}

@article{Hijano:2019qmi,
    author = "Hijano, Eliot",
    title = "{Flat space physics from AdS/CFT}",
    eprint = "1905.02729",
    archivePrefix = "arXiv",
    primaryClass = "hep-th",
    doi = "10.1007/JHEP07(2019)132",
    journal = "JHEP",
    volume = "07",
    pages = "132",
    year = "2019"
}

@article{Apolo:2020qjm,
	archiveprefix = {arXiv},
	author = {Apolo, Luis and Jiang, Hongliang and Song, Wei and Zhong, Yuan},
	doi = {10.1007/JHEP09(2020)033},
	eprint = {2006.10741},
	journal = {JHEP},
	pages = {033},
	primaryclass = {hep-th},
	title = {{Modular Hamiltonians in flat holography and (W)AdS/WCFT}},
	volume = {09},
	year = {2020}}

@article{Apolo:2020bld,
	archiveprefix = {arXiv},
	author = {Apolo, Luis and Jiang, Hongliang and Song, Wei and Zhong, Yuan},
	eprint = {2006.10740},
	month = {6},
	primaryclass = {hep-th},
	title = {{Swing surfaces and holographic entanglement beyond AdS/CFT}},
	year = {2020}}

@article{Chen:2020juc,
	archiveprefix = {arXiv},
	author = {Chen, Bin and Hao, Peng-xiang and Liu, Yan-jun},
	doi = {10.1103/PhysRevD.102.065016},
	eprint = {2006.04112},
	journal = {Phys. Rev. D},
	number = {6},
	pages = {065016},
	primaryclass = {hep-th},
	title = {{Supersymmetric Warped Conformal Field Theory}},
	volume = {102},
	year = {2020}}

@article{Chen:2020vvn,
    author = "Chen, Bin and Hao, Peng-Xiang and Liu, Reiko and Yu, Zhe-Fei",
    title = "{On Galilean conformal bootstrap}",
    eprint = "2011.11092",
    archivePrefix = "arXiv",
    primaryClass = "hep-th",
    doi = "10.1007/JHEP06(2021)112",
    journal = "JHEP",
    volume = "06",
    pages = "112",
    year = "2021"
}

@article{Hao:2021urq,
    author = "Hao, Peng-xiang and Song, Wei and Xie, Xianjin and Zhong, Yuan",
    title = "{BMS-invariant free scalar model}",
    eprint = "2111.04701",
    archivePrefix = "arXiv",
    primaryClass = "hep-th",
    doi = "10.1103/PhysRevD.105.125005",
    journal = "Phys. Rev. D",
    volume = "105",
    number = "12",
    pages = "125005",
    year = "2022"
}

@article{Fuentealba:2020zkf,
    author = "Fuentealba, Oscar and Gonz\'alez, Hern\'an A. and P\'erez, Alfredo and Tempo, David and Troncoso, Ricardo",
    title = "{Superconformal Bondi-Metzner-Sachs Algebra in Three Dimensions}",
    eprint = "2011.08197",
    archivePrefix = "arXiv",
    primaryClass = "hep-th",
    reportNumber = "CECS-PHY-20/04",
    doi = "10.1103/PhysRevLett.126.091602",
    journal = "Phys. Rev. Lett.",
    volume = "126",
    number = "9",
    pages = "091602",
    year = "2021"
}

@article{Frappat:1992bs,
    author = "Frappat, L. and Ragoucy, E. and Sorba, P.",
    title = "{W algebras and superalgebras from constrained WZW models: A Group theoretical classification}",
    eprint = "hep-th/9207102",
    archivePrefix = "arXiv",
    reportNumber = "ENSLAPP-AL-391-92",
    doi = "10.1007/BF02096881",
    journal = "Commun. Math. Phys.",
    volume = "157",
    pages = "499--548",
    year = "1993"
}

@article{Bouwknegt:1992wg,
    author = "Bouwknegt, Peter and Schoutens, Kareljan",
    title = "{W symmetry in conformal field theory}",
    eprint = "hep-th/9210010",
    archivePrefix = "arXiv",
    reportNumber = "CERN-TH-6583-92, ITPO-SB-92-23",
    doi = "10.1016/0370-1573(93)90111-P",
    journal = "Phys. Rept.",
    volume = "223",
    pages = "183--276",
    year = "1993"
}

@article{deBoer:1993iz,
    author = "de Boer, Jan and Tjin, Tjark",
    editor = "Bouwknegt, P. and Schoutens, K.",
    title = "{The Relation between quantum W algebras and Lie algebras}",
    eprint = "hep-th/9302006",
    archivePrefix = "arXiv",
    reportNumber = "THU-93-05A, THU-93-01A, ITFA-02-93",
    doi = "10.1007/BF02103279",
    journal = "Commun. Math. Phys.",
    volume = "160",
    pages = "317--332",
    year = "1994"
}

@article{2003Quantum,
  title={Quantum Reduction for Affine Superalgebras},
  author={ Kac, Victor  and  Roan, Shi Shyr  and  Wakimoto, Minoru },
  journal={Communications in Mathematical Physics},
  volume={241},
  number={2-3},
  pages={307-342},
  year={2003},
}

@article{Blumenhagen:1990jv,
    author = "Blumenhagen, R. and Flohr, M. and Kliem, A. and Nahm, W. and Recknagel, A. and Varnhagen, R.",
    editor = "Bouwknegt, P. and Schoutens, K.",
    title = "{W algebras with two and three generators}",
    reportNumber = "BONN-HE-90-05",
    doi = "10.1016/0550-3213(91)90624-7",
    journal = "Nucl. Phys. B",
    volume = "361",
    pages = "255--289",
    year = "1991"
}

@article{Chen:2022cpx,
    author = "Chen, Bin and Liu, Reiko",
    title = "{The Shadow Formalism of Galilean CFT$_2$}",
    eprint = "2203.10490",
    archivePrefix = "arXiv",
    primaryClass = "hep-th",
    month = "3",
    year = "2022"
}

@article{Chen:2022jhx,
    author = "Chen, Bin and Hao, Peng-xiang and Liu, Reiko and Yu, Zhe-fei",
    title = "{On Galilean Conformal Bootstrap II: $\xi=0$ sector}",
    eprint = "2207.01474",
    archivePrefix = "arXiv",
    primaryClass = "hep-th",
    month = "7",
    year = "2022"
}

@article{Bagchi:2016geg,
    author = "Bagchi, Arjun and Gary, Mirah and Zodinmawia",
    title = "{Bondi-Metzner-Sachs bootstrap}",
    eprint = "1612.01730",
    archivePrefix = "arXiv",
    primaryClass = "hep-th",
    doi = "10.1103/PhysRevD.96.025007",
    journal = "Phys. Rev. D",
    volume = "96",
    number = "2",
    pages = "025007",
    year = "2017"
}

@article{Bagchi:2017cpu,
    author = "Bagchi, Arjun and Gary, Mirah and Zodinmawia",
    title = "{The nuts and bolts of the BMS Bootstrap}",
    eprint = "1705.05890",
    archivePrefix = "arXiv",
    primaryClass = "hep-th",
    doi = "10.1088/1361-6382/aa8003",
    journal = "Class. Quant. Grav.",
    volume = "34",
    number = "17",
    pages = "174002",
    year = "2017"
}

@article{Chen:2021xkw,
    author = "Chen, Bin and Liu, Reiko and Zheng, Yu-fan",
    title = "{On Higher-dimensional Carrollian and Galilean Conformal Field Theories}",
    eprint = "2112.10514",
    archivePrefix = "arXiv",
    primaryClass = "hep-th",
    month = "12",
    year = "2021"
}

@article{Belavin:1984vu,
    author = "Belavin, A. A. and Polyakov, Alexander M. and Zamolodchikov, A. B.",
    editor = "Khalatnikov, I. M. and Mineev, V. P.",
    title = "{Infinite Conformal Symmetry in Two-Dimensional Quantum Field Theory}",
    reportNumber = "CERN-TH-3827",
    doi = "10.1016/0550-3213(84)90052-X",
    journal = "Nucl. Phys. B",
    volume = "241",
    pages = "333--380",
    year = "1984"
}

@article{Bagchi:2017cte,
    author = "Bagchi, Arjun and Banerjee, Aritra and Chakrabortty, Shankhadeep and Parekh, Pulastya",
    title = "{Inhomogeneous Tensionless Superstrings}",
    eprint = "1710.03482",
    archivePrefix = "arXiv",
    primaryClass = "hep-th",
    doi = "10.1007/JHEP02(2018)065",
    journal = "JHEP",
    volume = "02",
    pages = "065",
    year = "2018"
}

@article{Bagchi:2016yyf,
    author = "Bagchi, Arjun and Chakrabortty, Shankhadeep and Parekh, Pulastya",
    title = "{Tensionless Superstrings: View from the Worldsheet}",
    eprint = "1606.09628",
    archivePrefix = "arXiv",
    primaryClass = "hep-th",
    reportNumber = "MIT-CTP-4816",
    doi = "10.1007/JHEP10(2016)113",
    journal = "JHEP",
    volume = "10",
    pages = "113",
    year = "2016"
}

@article{Gamboa:1989px,
    author = "Gamboa, J. and Ramirez, C. and Ruiz-Altaba, M.",
    title = "{QUANTUM NULL (SUPER)STRINGS}",
    reportNumber = "CERN-TH-5367-89",
    doi = "10.1016/0370-2693(89)90578-9",
    journal = "Phys. Lett. B",
    volume = "225",
    pages = "335--339",
    year = "1989"
}

@article{Campoleoni:2016vsh,
    author = "Campoleoni, Andrea and Gonzalez, Hernan A. and Oblak, Blagoje and Riegler, Max",
    editor = "Brink, Lars and Henneaux, Marc and Vasiliev, Mikhail A.",
    title = "{BMS Modules in Three Dimensions}",
    eprint = "1603.03812",
    archivePrefix = "arXiv",
    primaryClass = "hep-th",
    doi = "10.1142/S0217751X16500688",
    journal = "Int. J. Mod. Phys. A",
    volume = "31",
    number = "12",
    pages = "1650068",
    year = "2016"
}

@article{Afshar:2013vka,
    author = "Afshar, Hamid and Bagchi, Arjun and Fareghbal, Reza and Grumiller, Daniel and Rosseel, Jan",
    title = "{Spin-3 Gravity in Three-Dimensional Flat Space}",
    eprint = "1307.4768",
    archivePrefix = "arXiv",
    primaryClass = "hep-th",
    reportNumber = "TUW-13-09",
    doi = "10.1103/PhysRevLett.111.121603",
    journal = "Phys. Rev. Lett.",
    volume = "111",
    number = "12",
    pages = "121603",
    year = "2013"
}

@article{Rasmussen:2017eus,
    author = "Rasmussen, Jorgen and Raymond, Christopher",
    title = "{Galilean contractions of $W$-algebras}",
    eprint = "1701.04437",
    archivePrefix = "arXiv",
    primaryClass = "hep-th",
    doi = "10.1016/j.nuclphysb.2017.07.006",
    journal = "Nucl. Phys. B",
    volume = "922",
    pages = "435--479",
    year = "2017"
}

@article{Creutzig:2013hma,
    author = "Creutzig, Thomas and Ridout, David",
    title = "{Logarithmic Conformal Field Theory: Beyond an Introduction}",
    eprint = "1303.0847",
    archivePrefix = "arXiv",
    primaryClass = "hep-th",
    doi = "10.1088/1751-8113/46/49/494006",
    journal = "J. Phys. A",
    volume = "46",
    pages = "4006",
    year = "2013"
}

@article{Hsieh:2020uwb,
    author = "Hsieh, Chang-Tse and Nakayama, Yu and Tachikawa, Yuji",
    title = "{Fermionic minimal models}",
    eprint = "2002.12283",
    archivePrefix = "arXiv",
    primaryClass = "cond-mat.str-el",
    reportNumber = "IPMU-20-0008, RUP-20-5",
    doi = "10.1103/PhysRevLett.126.195701",
    journal = "Phys. Rev. Lett.",
    volume = "126",
    number = "19",
    pages = "195701",
    year = "2021"
}

@book{DiFrancesco:1997nk,
    author = "Di Francesco, P. and Mathieu, P. and Senechal, D.",
    title = "{Conformal Field Theory}",
    doi = "10.1007/978-1-4612-2256-9",
    isbn = "978-0-387-94785-3, 978-1-4612-7475-9",
    publisher = "Springer-Verlag",
    address = "New York",
    series = "Graduate Texts in Contemporary Physics",
    year = "1997"
}

@article{Dotsenko:1984ad,
    author = "Dotsenko, V. S. and Fateev, V. A.",
    title = "{Four Point Correlation Functions and the Operator Algebra in the Two-Dimensional Conformal Invariant Theories with the Central Charge c \ensuremath{<} 1}",
    reportNumber = "NORDITA-84/22",
    doi = "10.1016/S0550-3213(85)80004-3",
    journal = "Nucl. Phys. B",
    volume = "251",
    pages = "691--734",
    year = "1985"
}

@article{Polchinski:1987dy,
    author = "Polchinski, Joseph",
    title = "{Scale and Conformal Invariance in Quantum Field Theory}",
    reportNumber = "UTTG-22-87",
    doi = "10.1016/0550-3213(88)90179-4",
    journal = "Nucl. Phys. B",
    volume = "303",
    pages = "226--236",
    year = "1988"
}

@article{Hofman:2011zj,
    author = "Hofman, Diego M. and Strominger, Andrew",
    title = "{Chiral Scale and Conformal Invariance in 2D Quantum Field Theory}",
    eprint = "1107.2917",
    archivePrefix = "arXiv",
    primaryClass = "hep-th",
    doi = "10.1103/PhysRevLett.107.161601",
    journal = "Phys. Rev. Lett.",
    volume = "107",
    pages = "161601",
    year = "2011"
}

@article{Chen:2019hbj,
    author = "Chen, Bin and Hao, Peng-Xiang and Yu, Zhe-Fei",
    title = "{2d Galilean Field Theories with Anisotropic Scaling}",
    eprint = "1906.03102",
    archivePrefix = "arXiv",
    primaryClass = "hep-th",
    doi = "10.1103/PhysRevD.101.066029",
    journal = "Phys. Rev. D",
    volume = "101",
    number = "6",
    pages = "066029",
    year = "2020"
}

@article{Bagchi:2019unf,
    author = "Bagchi, Arjun and Saha, Amartya and Zodinmawia",
    title = "{BMS Characters and Modular Invariance}",
    eprint = "1902.07066",
    archivePrefix = "arXiv",
    primaryClass = "hep-th",
    doi = "10.1007/JHEP07(2019)138",
    journal = "JHEP",
    volume = "07",
    pages = "138",
    year = "2019"
}

@article{Bagchi:2013qva,
    author = "Bagchi, Arjun and Basu, Rudranil",
    title = "{3D Flat Holography: Entropy and Logarithmic Corrections}",
    eprint = "1312.5748",
    archivePrefix = "arXiv",
    primaryClass = "hep-th",
    doi = "10.1007/JHEP03(2014)020",
    journal = "JHEP",
    volume = "03",
    pages = "020",
    year = "2014"
}

@article{Bagchi:2018wsn,
    author = "Bagchi, Arjun and Banerjee, Aritra and Chakrabortty, Shankhadeep and Parekh, Pulastya",
    title = "{Exotic Origins of Tensionless Superstrings}",
    eprint = "1811.10877",
    archivePrefix = "arXiv",
    primaryClass = "hep-th",
    doi = "10.1016/j.physletb.2019.135139",
    journal = "Phys. Lett. B",
    volume = "801",
    pages = "135139",
    year = "2020"
}

@article{Bagchi:2009pe,
    author = "Bagchi, Arjun and Gopakumar, Rajesh and Mandal, Ipsita and Miwa, Akitsugu",
    title = "{GCA in 2d}",
    eprint = "0912.1090",
    archivePrefix = "arXiv",
    primaryClass = "hep-th",
    reportNumber = "HRI-ST-0923",
    doi = "10.1007/JHEP08(2010)004",
    journal = "JHEP",
    volume = "08",
    pages = "004",
    year = "2010"
}

@article{Haco:2017ekf,
    author = "Haco, Sasha J. and Hawking, Stephen W. and Perry, Malcolm J. and Bourjaily, Jacob L.",
    title = "{The Conformal BMS Group}",
    eprint = "1701.08110",
    archivePrefix = "arXiv",
    primaryClass = "hep-th",
    doi = "10.1007/JHEP11(2017)012",
    journal = "JHEP",
    volume = "11",
    pages = "012",
    year = "2017"
}

@inproceedings{Ginsparg:1988ui,
    author = "Ginsparg, Paul H.",
    title = "{APPLIED CONFORMAL FIELD THEORY}",
    booktitle = "{Les Houches Summer School in Theoretical Physics: Fields, Strings, Critical Phenomena}",
    eprint = "hep-th/9108028",
    archivePrefix = "arXiv",
    reportNumber = "HUTP-88-A054",
    month = "9",
    year = "1988"
}

@article{Hao:2022xhq,
    author = "Hao, Peng-Xiang and Song, Wei and Xiao, Zehua and Xie, Xianjin",
    title = "{A BMS-invariant free fermion model}",
    eprint = "2211.06927",
    archivePrefix = "arXiv",
    primaryClass = "hep-th",
    month = "11",
    year = "2022"
}

@article{Fareghbal:2013ifa,
    author = "Fareghbal, Reza and Naseh, Ali",
    title = "{Flat-Space Energy-Momentum Tensor from BMS/GCA Correspondence}",
    eprint = "1312.2109",
    archivePrefix = "arXiv",
    primaryClass = "hep-th",
    doi = "10.1007/JHEP03(2014)005",
    journal = "JHEP",
    volume = "03",
    pages = "005",
    year = "2014"
}

@article{Saha:2022gjw,
    author = "Saha, Amartya",
    title = "{Intrinsic Approach to $1+1$D Carrollian Conformal Field Theory}",
    eprint = "2207.11684",
    archivePrefix = "arXiv",
    primaryClass = "hep-th",
    month = "7",
    year = "2022"
}

@ARTICLE{2022CMaPh.390...33F,
       author = {{Fasquel}, Justine},
        title = "{Rationality of the Exceptional W -Algebras W$_{k}$(sp$_{4}$,f$_{subreg}$) Associated with Subregular Nilpotent Elements of sp$_{4}$}",
      journal = {Communications in Mathematical Physics},
         year = 2022,
        month = feb,
       volume = {390},
       number = {1},
        pages = {33-65},
          doi = {10.1007/s00220-021-04294-6},
archivePrefix = {arXiv},
       eprint = {2009.09513},
 primaryClass = {math.RT},
       adsurl = {https://ui.adsabs.harvard.edu/abs/2022CMaPh.390...33F}
}

@article{Bagchi:2009my,
    author = "Bagchi, Arjun and Gopakumar, Rajesh",
    title = "{Galilean Conformal Algebras and AdS/CFT}",
    eprint = "0902.1385",
    archivePrefix = "arXiv",
    primaryClass = "hep-th",
    doi = "10.1088/1126-6708/2009/07/037",
    journal = "JHEP",
    volume = "07",
    pages = "037",
    year = "2009"
}

@article{Bagchi:2023dzx,
    author = "Bagchi, Arjun and Chatterjee, Ritankar and Kaushik, Rishabh and Saha, Amartya and Sarkar, Debmalya",
    title = "{Non-Lorentzian Ka\v{c}-Moody algebras}",
    eprint = "2301.04686",
    archivePrefix = "arXiv",
    primaryClass = "hep-th",
    doi = "10.1007/JHEP03(2023)041",
    journal = "JHEP",
    volume = "03",
    pages = "041",
    year = "2023"
}

@article{Grumiller:2026kgx,
    author = "Grumiller, Daniel and Lovrekovic, Iva",
    title = "{Unitary quantum conformal Bondi-Metzner-Sachs field theories}",
    eprint = "2609.07828",
    archivePrefix = "arXiv",
    primaryClass = "hep-th",
    reportNumber = "TUW-26-06",
    month = "9",
    year = "2026"
}
\end{refcontext}

\end{document}